\documentclass[aps,prl,twocolumn,amsmath,amssymb,floatfix]{revtex4-2}  

\usepackage[outline]{contour}
\usepackage{color}
\usepackage{bbm}
\usepackage{graphicx}
\usepackage{dcolumn}
\usepackage[utf8]{inputenc}
\usepackage{graphicx}
\usepackage{epstopdf}
\usepackage{amsthm}
\usepackage{amsmath}
\usepackage{empheq}
\usepackage{eufrak}
\usepackage{bbm}
\usepackage{braket}
\usepackage{amssymb}
\usepackage{mathtools}
\usepackage[usenames,dvipsnames]{xcolor}
\usepackage{latexsym}
\usepackage[colorlinks=true,citecolor=Cerulean,linkcolor=RubineRed,urlcolor=Cerulean]{hyperref}
\usepackage{cleveref}

\newcommand{\opt}{\textnormal{opt}}
\newcommand{\var}{\textnormal{var}}

\newcommand{\tr}[1]{\operatorname{\textnormal{Tr}}\left( {#1} \right)}

\newcommand{\info}{F}
\newcommand{\cl}{\textnormal{isolated}}  \newcommand{\open}{\textnormal{open}}

\begin{document}

\title{
Noise-enhanced quantum clocks and global field sensors 
}
 
\author{Luis Pedro Garc\'ia-Pintos}
\email{lpgp@lanl.gov}
\affiliation{Quantum and Condensed Matter Physics Group (T-4), Theoretical Division, Los Alamos National Laboratory, Los Alamos, New Mexico 87545, USA}

\begin{abstract}
I show that incoherent dynamics can lead to metrological advantages in quantum sensing. The results rely on the fact that incoherent dynamics lead to an additive contribution to the quantum Fisher information about time. Such an additive contribution can lead to a decrease in the error of optimal estimation protocols, as implied by the quantum Cram\'er-Rao bound. I characterize regimes in which the estimation of a time interval or a frequency is enhanced by noise, thus identifying cases where incoherent dynamics serve as a metrological resource. 
I illustrate with protocols that display improved sensing of time intervals or global fields by qubit and photonic sensor networks. 
\end{abstract}

\maketitle
Quantum sensors are systems that rely on the rules of quantum physics to estimate a quantity of interest, e.g., to measure a magnetic field or a time interval. In certain regimes, quantum sensors surpass the precision of classical sensors. Quantum sensing and metrology constitute leading applications of presently available quantum technologies. For example, LIGO exploits the phenomenon of quantum squeezing for gravitational wave detection~\cite{PhysRevLett.68.3020, LIGO}.

Other models of quantum sensors leverage quantum coherence and entanglement (a stronger-than-classical form of correlation) to sense. For instance, consider estimating a magnetic field $\omega$. A coherent sensor network of $N$ non-interacting entangled spins can estimate $\omega$ with a precision that scales as $1/N$~\cite{Bollinger1996,GiovannettiPRL2006}. Meanwhile, uncorrelated spins yield a precision that scales as $1/\sqrt{N}$: a $\sqrt{N}$ system-size advantage in estimating the field. In this way, to varying degrees, quantum coherence and correlations, or more generally non-classicality, can be thought of as resources for quantum metrology~\cite{GiovannettiPRL2006, tan2019quantum, LostaglioPRL2020, PhysRevResearch.5.033228, brady2024correlated}. 

Coherence and entanglement are fragile resources. Under the influence of uncontrolled noises or the interaction with an environment, coherence and entanglement typically vanish rapidly~\cite{Schlosshauer}. 
Substantial literature is thus devoted to the detrimental effects of decoherence and noise on sensing and metrology. For overviews, see Refs.~\cite{precisionlimitsopen,Datta2025}. As a result of noise, quantum advantages are typically negated~\cite{escher2011quantum, demkowicz2012NatComm, tsang2013quantum, PhysRevX.7.041009}. This fact has motivated detailed characterizations of the influence of noise on quantum sensors~\cite{escher2011general, PhysRevLett.112.120405, Zhou, len2022quantum, PhysRevA.111.L020403}, identifying metrological techniques that can yield quantum advantages despite the presence of particular sources of noise~\cite{PhysRevA.84.012103, Noisymetro, PRXQuantum.3.030330, PhysRevLett.133.080801}, or applying error correction protocols to recover the metrological power of ideal isolated quantum sensors~\cite{QECPE1, PhysRevLett.133.190801, Zhou, QECPE2}. An ultimate goal is identifying scenarios where quantum sensors can surpass classical ones, possibly under the effect of noise, for instance by exploiting many-body effects~\cite{adc, Beau:17}, superradiance~\cite{PRXQuantum.3.030330}, long-range interactions~\cite{PhysRevResearch.4.013133}, non-linear state dynamics~\cite{deffner}, or criticality~\cite{PhysRevA.78.042105, PhysRevLett.121.020402, PRXQuantum.3.010354, yu2024criticality, PhysRevA.109.L050601}. Notably, Ref.~\cite{adc} proves that nonlocal many-body Hamiltonians and Lindbladians that act on many systems can lead to improved parameter sensitivity, and  Ref.~\cite{wang2024exponential} identifies entanglement advantages in estimating Lindbladian decay rates. Most related to this work, Ref.~\cite{PhysRevLett.133.090801} shows that non-Hermitian sensors can benefit from noise, and  Ref.~\cite{PhysRevLett.133.190801} shows protocols that exploit error correction to leverage noise for precision improvements when estimating Hamiltonian parameters.

In this letter, I identify regimes in which noise can enhance the metrological power of a sensor network. The approach is conceptually related to engineered noise and dissipation~\cite{verstraete2009quantum, PhysRevA.83.012304, PhysRevLett.118.140403, Harrington_2022, Sannia_2024, martinez2024quantum}, which aims to leverage open dynamics to one's favor. I prove a noise-induced sensitivity improvement over the best sensitivity of an isolated quantum sensor when estimating time intervals or global Hamiltonian parameters.

\vspace{12pt}
\noindent \textbf{Background: minimum estimation errors and metrology with isolated quantum sensors}    
\vspace{2pt}

Consider a quantum sensor whose state $\rho$ depends on a parameter $\lambda$. The quantum Cram\'er-Rao bound limits the minimum uncertainty with which $\lambda$ can be estimated from measurements on $\rho$. For any observable $O$, the quantum Cram\'er-Rao bound states that~\cite{helstrom1969quantum}
\begin{align}
\label{eq:CRbound}
\var\big( O \big) \frac{1}{\left| \frac{\partial}{\partial \lambda} \langle O \rangle \right|^2} \geq \frac{1}{\info(\lambda)},
\end{align}
where $\langle O \rangle = \tr{O \rho}$ and $\var(O) = \langle O^2 \rangle - \langle O \rangle^2$ are the operator's expectation value and variance, respectively. The \emph{Fisher information} about the parameter $\lambda$ is $\info(\lambda) \nobreak\coloneqq\nobreak \tr{\rho \mathcal{L}^2(\lambda)}$, where the symmetric logarithmic derivative $\mathcal{L(\lambda)}$ is implicitly defined by $\partial \rho/\partial \lambda = \{\rho, \mathcal{L}(\lambda)\}/2$~\cite{liu2019quantum}. The bound~\eqref{eq:CRbound} is saturable by a suitable choice of $O$. (However, the observable that saturates the bound may depend on the parameter to be estimated~\cite{paris2009quantum}.)

By propagation of error, the uncertainty in any estimator $\hat \lambda$ of $\lambda$ is $\var(\hat \lambda) =  \left| \frac{\partial}{\partial \lambda} \langle O \rangle \right|^{-2}\var\big( O \big)$~\cite{GiovannettiPRL2006}. Thus, Eq.~\eqref{eq:CRbound} bounds the estimation error of the parameter $\lambda$. Moreover, since the bound is saturable, the quantum Fisher information $F(\lambda)$ about a parameter determines the minimum estimation error. Often, the right-hand side of the Cram\'er-Rao bound~\eqref{eq:CRbound} contains a prefactor $1/M$ to account for the statistical error when performing $M$ measurements on $M$ copies of the state $\rho$~\cite{helstrom1969quantum}. 

To find the minimum estimation error of a parameter, one can thus focus on the quantum Fisher information. 
Consider an ideal quantum sensor that evolves unitarily under a Hamiltonian $H$, unaffected by noise or environments. The sensor's (pure) state $\rho_t$ at time $t$ satisfies
\begin{align}
\label{eq:closed}
\frac{d \rho_t}{dt} = -i[H,\rho_t].
\end{align} 
If the Hamiltonian depends on the $\lambda$ that one wishes to estimate, $\rho_t$ will generally acquire a dependence on $\lambda$  too. 

For example, let the goal be to estimate $\lambda \equiv t$ (i.e., we want to use the sensor as a clock). The Fisher information about $t$ is $\info_\cl(t) = 4 \var (H)$, where $\var (H) = { \tr{\rho_t H^2} - (\tr{\rho_t H})^2}$ is the sensor's energy variance~\cite{BraunsteinCaves1994}. The Cram\'er-Rao bound~\eqref{eq:CRbound} then implies that an optimal estimation protocol (one that saturates the bound via an estimator $\hat{t}_\opt$) determines $t$ with an uncertainty 
\begin{align}
\label{eq:uncertaintytime}
    \var_\cl \big(\hat t_{\opt} \big)\nobreak =\nobreak \frac{1}{ \info_\cl(t)} \nobreak=\nobreak \frac{1}{4 \var (H)}.
\end{align}

Alternatively, consider the estimation of a global parameter $\omega$ in $H$. For example, $\omega$ could be a magnetic field that couples to $N$ independent qubits or the frequency of independent harmonic oscillators. The minimum uncertainty in $\omega$ has a similar expression to Eq.~\eqref{eq:uncertaintytime}:  
\begin{align}
\label{eq:uncertaintyfrequency}
    \var_\cl \big(\hat \omega_\opt \big)\nobreak =\nobreak \frac{1}{ \info_\cl(\omega)} \nobreak=\nobreak \frac{\omega^2}{4  \var (H) t^2}.
\end{align}
Thus, the best metrological states for time and frequency estimation maximize the sensor's energy variance.

To illustrate the metrological power of such quantum sensors, consider an $N$-qubit sensor network with a Hamiltonian $H = \omega \sum_{l = 1}^{N} \sigma_l^z/2$, where $\sigma_l^z$ is the Pauli $Z$-matrix of qubit $l$, with eigenstates $\ket{0}_l$ and $\ket{1}_l$. The sensor network starts in a correlated cat state
\begin{align}
\label{eq:Cat}
\ket{\psi} = \frac{1}{\sqrt{2}} \big( \ket{E_0} + \ket{E_1} \big),
\end{align}
where $\ket{E_0}$ and $\ket{E_1}$ are eigenstates of the Hamiltonian. If $\ket{E_0} \coloneqq \otimes_j \ket{0}_j$ and  $\ket{E_1} \coloneqq \otimes_j \ket{1}_j$ (i.e., $\ket{\psi}$ is a GHZ state), the energy variance is $\var(H) = N^2\omega^2/4$. Then, the sensor's precision in estimating time and frequency improve with its size as $\var_\cl \big(\hat t_\opt \big)\nobreak =\nobreak \frac{1}{ \omega^2 N^2 }$ and $\var_\cl \big(\hat \omega_\opt \big)\nobreak =\nobreak \frac{\omega^2}{ t^2 N^2 }$~\cite{Bollinger1996, GiovannettiPRL2006}. The same precision is obtained by a two-mode photonic sensor with a Hamiltonian $H = \omega a^\dag a /2 + \omega b^\dag b /2$ in a NOON state $\ket{\psi}$, with $\ket{E_1} = \ket{0}_A\otimes\ket{N}_B$ and $ \ket{E_0} = \ket{N}_A\otimes\ket{0}_B$~\cite{NOON1, NOON2}.  

In these paradigmatic examples, the sensitivity of the sensors crucially depends on the quantum coherence in the initial cat state~\eqref{eq:Cat}. For instance, if the sensors were to reach the energy decohered counterpart $(\ket{E_0}\!\bra{E_0} + \ket{E_1}\!\bra{E_1} )/2$ of the cat state~\eqref{eq:Cat}, the quantum Fisher information about time or frequency would be null, making for a lousy sensor. Generally, decoherence and noisy quantum dynamics decrease the quantum Fisher information of a sensor network, hindering its sensitivity~\cite{escher2011quantum, precisionlimitsopen, tsang2013quantum, Datta2025}. However, this is not always the case. Next, I show that tailored open dynamics can enhance the metrological power of a quantum sensor.

\vspace{12pt}
\noindent \textbf{Enhanced metrology with open quantum sensors}
\vspace{2pt}

Under non-unitary transformations, two positive and additive terms contribute to the quantum Fisher information: $F_\open(\lambda) \nobreak=\nobreak F_{\textnormal{unitary}}(\lambda) \nobreak+\nobreak F_{\textnormal{incoherent}}(\lambda)$~\cite{paris2009quantum, PhysRevA.90.022111}. The first term includes all contributions from coherent state transformations (e.g., those generated by $e^{-i \lambda K}$ for a Hermitian $K$), which can typically be identified as responsible for quantum advantage in metrological settings. The second term, $F_{\textnormal{incoherent}}$, is directly related to incoherent, possibly entropy-changing transformations. These occur, for instance, due to noisy non-unitary processes or environments affecting the system.
 
Since the saturability of the Cram\'er-Rao bound~\eqref{eq:CRbound} links higher Fisher information to better estimation precision, the fact that $F_{\textnormal{incoherent}}$ is positive suggests that regimes may exist where a sensor benefits from incoherent dynamics. 
As I noted before, however, incoherent dynamics tends to decrease a quantum sensor's precision~\cite{escher2011quantum, demkowicz2012NatComm, PhysRevX.7.041009}. Thus, to yield considerable enhancements in the sensing precision, one may need incoherent dynamics tailored to enhance sensitivity to a parameter.

Motivated by such realizations, I consider a sensor that evolves isolated [according to Eq.~\eqref{eq:closed}] until a time $t_0$, at which time the system becomes open. The dynamics for $t > t_0$ is given by  
\begin{align}
\label{eq:open}
\frac{d\rho_t}{dt} = -i [H,\rho_t] - \gamma_t \big[ L,[L,\rho_t] \big].
\end{align}
I assume that $L$ is a Hermitian Lindblad operator such that $[H,L] = 0$. Equation~\eqref{eq:open} describes the dynamics of an open quantum system losing coherence in the eigenbasis of $L$ at a (possibly time-dependent) rate $\gamma_t$. Such decoherence can occur due to the interactions with an environment~\cite{Schlosshauer} or due to dynamics with a stochastic Hamiltonian $H + \xi_t L$, where $\xi_t$ is a zero-mean white noise process of variance $\gamma_t$~\cite{PhysRevLett.118.140403}. Noise and decoherence described by $L$ and a rate $\gamma_t$ have the same effect on the system's dynamics.
  
\vspace{7pt}

\noindent\emph{Estimating time.---}For a quantum sensor evolving under Eq.~\eqref{eq:open} for $t > t_0$, I prove in Sec.~\ref{app:QFItime} of the Appendix that the quantum Fisher information about time satisfies
\begin{align}
\label{eq:BoundFtime}
    F_\open(t) \geq \left\| [H,\rho_t] \right\|_2^2 + \gamma_t^2 \left\| [L,[L,\rho_t]] \right\|_2^2,
\end{align}
where $\|A\|_2 \coloneqq \sqrt{\tr{A A^\dag} }$ is the Hilbert-Schmidt operator norm. Recall that optimal estimators saturate the Cram\'er-Rao bound. Thus, the lower bound~\eqref{eq:BoundFtime} on the quantum Fisher information sets an upper bound on the precision in estimating $t$ by optimal procedures. That is, $t$ can be estimated with an error $\var (\hat t_\opt) \nobreak \leq \nobreak 1/\big(\left\| [H,\rho_t] \right\|_2^2 + \gamma_t^2 \left\| [L,[L,\rho_t]] \right\|_2^2 \big)$.

Equation~\eqref{eq:BoundFtime} further suggests that noise may enhance time estimation: for a given $\rho_t$, the term proportional to the dephasing rate $\gamma_t$ increases the bound on the quantum Fisher information about time. 
However, the state loses coherence as dephasing acts, which leads to a decrease of both terms in the right-hand side of Eq.~\eqref{eq:BoundFtime}. Thus, if incoherent dynamics can enhance a quantum sensor, it should act for short periods of time (hence the assumption that dephasing acts at times $t > t_0$). 

Let the sensor network start in the cat state~\eqref{eq:Cat}, where $\ket{E_0}$ and $\ket{E_1}$ are joint eigenvectors of $L$ and $H$. For $t > t_0$, the quantum Fisher information about time is 
\begin{align}
\label{eq:FtimeOpen}
\info_\open(t) =& e^{- 2 \,  (\delta L)^2 \int_{t_0}^t \gamma_s ds}  \\  
& \qquad \left( (\delta E)^2  + \frac{   \gamma_t^2 \, (\delta L)^4  }{1 - e^{-2 \, (\delta L)^2 \int_{t_0}^t \gamma_s ds} }  \right),  \nonumber 
\end{align}
where $\delta E \coloneqq H \ket{E_1} - H \ket{E_0}$ and $\delta L \coloneqq L \ket{E_1} - L \ket{E_0}$ are the eigenvalue differences of $H$ and $L$, are respectively (see proof in Sec.~\ref{app:QFItime} of the Appendix). The exponential prefactor in Eq.~\eqref{eq:FtimeOpen} worsens the optimal precision of the quantum sensor as $t$ grows. However, there are regimes where dephasing enhances the estimation error.

 For comparison, the Fisher information about time of a sensor that evolves under a Hamiltonian $H$, isolated from noise or the environment, is $F_\cl(t) = (\delta E)^2$. [See Eq.~\eqref{eq:uncertaintytime} and note that the sensor's energy variance is $\var(H) = (\delta E)^2/4$.] Then, a quantum sensor can use noise as a metrological resource to estimate $t$ whenever
\begin{align}
\label{eq:NoiseAdvantage}
  1 <& \frac{\var_\cl(\hat t_\opt)}{\var_\open(\hat t_\opt)} = \frac{\info_\open(t)}{\info_\cl(t)}  \\
  &=  e^{- 2 \,  (\delta L)^2 \int_{t_0}^t \gamma_s ds} \left(  1   + \frac{   \gamma_t^2 \, (\delta L)^4/(\delta E)^2  }{1 - e^{-2 \, (\delta L)^2 \int_{t_0}^t \gamma_s ds} }  \right) \nonumber
\end{align}
holds. Equation~\eqref{eq:NoiseAdvantage} identifies regimes where the precision in estimating time increases by open system dynamics.  

To illustrate concrete noise-enhanced cases, consider an example where the dephasing rate is ramped up linearly in time, with $\gamma_t = \dot \gamma (t-t_0)$, where $\dot \gamma$ is constant (this model is more physical than assuming an instantaneous quench of the dephasing rate). If the time interval over which dephasing acts is $t-t_0 = \frac{\sqrt{\ln(2)}}{\sqrt{\dot \gamma} \delta L}$, the Fisher information is
\begin{align}
\label{FisherTimeCat}
    F_\open(t) = F_\cl(t) \left( \frac{1}{2} +  \frac{ \dot \gamma (\delta L)^2}{(\delta E)^2} \right). 
\end{align}
In such a case, Eq.~\eqref{eq:NoiseAdvantage} holds if $\dot \gamma (\delta L)^2 / (\delta E)^2 > \frac{1}{2}$.
The minimum precision with which one can determine $t$ thus satisfies
\begin{align}
\label{eq:CatTime}
 \var_\open(\hat t_\opt) = \frac{\var_\cl(\hat t_\opt)}{  \dot \gamma (\delta L)^2/(\delta E)^2 + 1/2 }.
\end{align}
The sensor improves as $\dot \gamma (\delta L)^2 / (\delta E)^2$ increases. 

In the above example, an open-system advantage requires sufficient control to let the system evolve for a time interval $t-t_0 = \sqrt{\tfrac{\ln(2)}{\dot \gamma (\delta L)^2}}$, which much satisfy $t-t_0 < \sqrt{2\ln(2)}/\delta E$ for condition~\eqref{eq:NoiseAdvantage} to hold. Such a technique could be helpful in high precision estimation of $t$ in scenarios where (i) one can enforce open system dynamics~\eqref{eq:open} for $t>t_0$, and (ii) one has access to a precise enough stopwatch that can determine a time interval $t-t_0 \sim 1/\delta E$ (i.e., the stopwatch must be as precise as the original isolated sensor is). Such a protocol could noise-enhance the accuracy of a clock (how precisely it can determine $t$), assuming access to a high-resolution stopwatch that can determine a much smaller $\delta t = t-t_0$~\cite{PhysRevX.7.031022, PhysRevLett.131.220201}.

The case described above serves as proof-of-principle that noisy dynamics can lead to metrological advantages in sensing a time interval $t$. Next, I explore using incoherent dynamics to enhance frequency estimation.

\vspace{7pt}

\noindent \emph{Estimating frequencies.---}Consider a quantum sensor affected by decoherence in energy, i.e., with $L = H$ in Eq.~\eqref{eq:open}. Such decoherence occurs, in particular, due to the use of an imperfect clock to determine the duration of a dynamical process~\cite{EgusquizaClocks1999, GambiniClocks2004, HuberClocks2023}. The quantum Fisher information about a global parameter $\omega$ in the Hamiltonian $H = \omega h$, where $h$ is Hermitian, satisfies (see Sec.~\ref{app:QFIfrequency} of the Appendix)
\begin{align}
\label{eq:BoundFfrequency}
    F_\open(\omega) &\geq \frac{t^2}{\omega^2} \left\| [H,\rho_t]  \right\|_2^2  +  \frac{4 \left( \int_{t_0}^t \gamma_s ds\right)^2 }{\omega^2} \left\| [H,[H,\rho_t]] \right\|_2^2.
\end{align}
In a similar way to Eq.~\eqref{eq:BoundFtime} for time estimation, Eq.~\eqref{eq:BoundFfrequency} suggests that dephasing may enhance the estimation of $\omega$ in certain regimes.

  \begin{figure}[!htbp]
      \centering
     $\var_\cl(\hat \omega_\opt) /  \var_\open(\hat \omega_\opt)$ as a function of $\omega$ and $\gamma$
      \includegraphics[width = 0.5\textwidth]{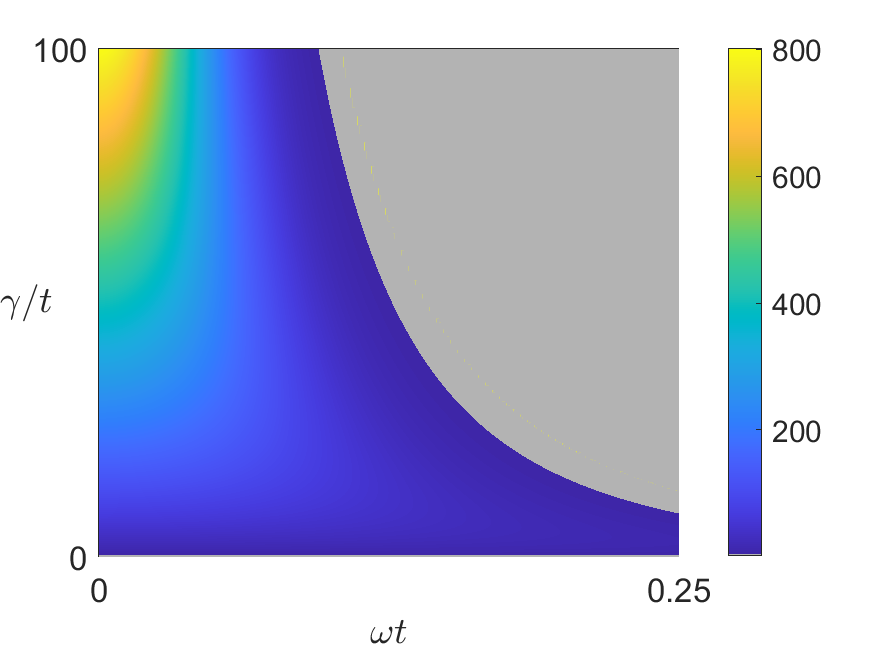}
      \caption{ \textbf{Noise-enhanced quantum sensing of frequencies.} 
Heatmap comparing the estimation error $\var_\cl(\hat \omega_\opt)$ of isolated quantum sensors versus the error $\var_\open(\hat \omega_\opt)$ of noise-enhanced quantum sensors when estimating a global Hamiltonian parameter $\omega$. When $ \frac{\var_\cl(\hat \omega_\opt)}{ \var_\open(\hat \omega_\opt)} > 1$, the noise-enhanced sensor is more precise than the isolated one. The isolated sensor evolves under $H$ for a time $t$, while the open sensor evolves under $H$ affected by energy decoherence at a rate $\gamma$ for a time $t$. The sensor starts in a cat state $\ket{\psi} = (\ket{E_0} + \ket{E_1})/\sqrt{2}$, where $\ket{E_0}$ and $\ket{E_1}$ are energy eigenstates with energy difference $\delta E$.  For the plot, I assume $\delta E = 2\omega$; the energy difference of a 2-spin sensor with Hamiltonian $H = \omega \sum_{l = 1}^{N} \sigma_l^z/2$ in a GHZ state, or the energy difference of a photonic sensor with $H = \omega a^\dag a /2 + \omega b^\dag b /2$ in a 2-excitation NOON state. The gray region corresponds to $ \frac{\var_\cl(\hat \omega_\opt)}{ \var_\open(\hat \omega_\opt)} < 1$, where noise hinders the estimation precision. In all other regions, where $ \frac{\var_\cl(\hat \omega_\opt)}{ \var_\open(\hat \omega_\opt)} > 1$, noise serves as a resource that improves the precision of the quantum sensor. At small $\omega t$ and large $\gamma/t$, the precision improves by three orders of magnitude.}
      \label{figure}
  \end{figure}

For a sensor initialized in state~\eqref{eq:Cat} that evolves according to Eqs.~\eqref{eq:closed} and~\eqref{eq:open} for time $t < t_0$ and  $t>t_0$, respectively, the quantum Fisher information about $\omega$ is (see Sec.~\ref{app:QFIfrequency} of the Appendix)
 \begin{align}
 \label{eq:FfrequencyOpen}
F_\open(\omega) &=  \frac{(\delta E)^2}{\omega^2} e^{- 2 (\delta E)^2 \int_{t_0}^t \gamma_s ds }   \\
&\qquad \left(   4 (\delta E)^2    \frac{\left( \int_{t_0}^t \gamma_s  ds \right)^2}{\left( 1- e^{-2 (\delta E)^2 \int_{t_0}^t \gamma_s ds } \right)}   +  t^2   \right). \nonumber 
\end{align}

In comparison, the quantum Fisher information about $\omega$ of a sensor that evolves isolated from noise or the environment is $F_\cl(\omega) = (\delta E)^2 t^2 / \omega^2$ [see Eq.~\eqref{eq:uncertaintyfrequency}]. Thus, the following condition identifies regimes where the precision in estimating $\omega$ increases by open system dynamics:
\begin{align}
\label{eq:NoiseAdvantageFreq}
  1 <& \frac{\var_\cl(\hat \omega_\opt)}{\var_\open(\hat \omega_\opt)} = \frac{\info_\open(\omega)}{\info_\cl(\omega)}  \\
  &=  e^{- 2 \,  (\delta E)^2 \int_{t_0}^t \gamma_s ds} \left(       \frac{ 4 (\delta E)^2 \left( \int_{t_0}^t \gamma_s  ds \right)^2}{t^2 \left( 1- e^{-2 (\delta E)^2 \int_{t_0}^t \gamma_s ds } \right)}   +  1  \right). \nonumber
\end{align}

For instance, consider a system affected by energy decoherence at a constant rate $\gamma$ that acts throughout the whole sensing protocol (i.e., $t_0 = 0$). If the sensing time is $t = \frac{ \ln(2) }{ 2 \gamma (\delta E)^2}$, the quantum Fisher information about $\omega$ is
\begin{align}
\label{eq:CatFrequency}
    F_\open(\omega) = F_\cl(\omega) \left(   \frac{1}{2} + 4 \gamma^2 (\delta E)^2 \right).
\end{align}
The condition~\eqref{eq:NoiseAdvantageFreq} for a noise-induced sensing advantage holds if $4 \gamma^2 (\delta E)^2 > 1/2$. For this to be the case, the sensing time must satisfy $t <  \sqrt{2} \ln (2) / \delta E$. Enhancing a quantum sensor by energy decoherence could thus be of interest for high-precision sensing of small $\omega$'s (e.g., due to spurious global fields), when the energy difference $\delta E$ is small.  
Figure~\ref{figure} explores more general regimes where noise enhances a quantum sensor's precision to estimate $\omega$.

As mentioned in the introduction, precision improvements of a Hamiltonian parameter were also observed in Ref.~\cite{PhysRevLett.133.190801} by exploiting quantum error correction. The noise-induced metrological advantages introduced in this letter [Eqs.~\eqref{eq:NoiseAdvantage} and~\eqref{eq:NoiseAdvantageFreq}] manifest in simple protocols that do not require error correction or error mitigation techniques. I show this next.

\vspace{12pt}
\noindent \textbf{Optimal estimators that benefit from noise}
\vspace{2pt}

Consider the canonical example described after Eq.~\eqref{eq:uncertaintyfrequency}, where an $N$-qubit sensor network with a Hamiltonian $H = \omega \sum_{l = 1}^{N} \sigma_l^z/2$ is used to estimate time or the global field $\omega$. Here, I assume the sensor network evolves under Hamiltonian $H$ until $t_0$, at which point energy decoherence starts to act. If the $N$ qubits are initialized in a GHZ state, $\delta E = N \omega$.

Let $O \coloneqq \bigotimes_{l = 1}^N \sigma_l^x$, where $\sigma_l^x$ is the Pauli $X$ matrix of qubit $l$~\cite{GiovannettiPRL2006}. The observable's expectation value is $\langle O \rangle_t = \cos( \delta E t ) e^{-\delta E^2 \int_{t_0}^t \gamma_s ds}$. Then, an estimator $\hat t$ of $t$ constructed from $\langle O \rangle$ has an error given by
\begin{align}
\label{eq:CatTimeExample}
     &\var( \hat t_\opt  ) = \frac{(\Delta O)^2}{\left| \frac{\partial \langle O \rangle_t}{\partial t}  \right|^2}  \\
     &= \var_\cl(\hat t_\opt) \left( \frac{ 2 - \cos^2( \delta E t )  }{ \left| \sin (\delta E t) + \sqrt{\ln(2) \dot \gamma }  \cos(\delta E t) \right|^2 } \right)   .\nonumber
\end{align}
(See details in Sec.~\ref{app:optimalestimators}  of the Appendix.) At times when $t \delta E$ is an integer multiple of $\pi$, Eq.~\eqref{eq:CatTimeExample} decreases with $\gamma^{-1}$, as does the optimal precision implied by Eq.~\eqref{eq:CatTime}.

Measuring $\langle O \rangle$ to estimate $t$ saturates the noise-enhancement implied by Eqs.~\eqref{FisherTimeCat} and~\eqref{eq:CatTime}. Due to the oscillatory nature of $\langle O \rangle$, estimating $t$ from it requires some broad prior information about $t$, typically leading to worse constant prefactors in the estimation error~\cite{PhysRevLett.124.030501, PhysRevA.102.042613}. I show in Sec.~\ref{app:optimalestimators} of the Appendix that $\langle O \rangle$ can also estimate $\omega$ with a precision that saturates the one implied by Eq.~\eqref{eq:CatFrequency}.

Finally, similar calculations show that photonic quantum sensors can also benefit from incoherent noise to sense time or frequencies. Consider a two-mode photonic sensor with 
a Hamiltonian $H = \omega a^\dag a /2 + \omega b^\dag b /2$. The sensor is initialized in a NOON state, given by Eq.~\eqref{eq:Cat} with $\ket{E_1} = \ket{0}_A\otimes\ket{N}_B \equiv \ket{0,N}$ and $ \ket{E_0} = \ket{N}_A\otimes\ket{0}_B \equiv \ket{N,0}$~\cite{NOON1, NOON2}. The observable $O \coloneqq \ket{0,N}\!\bra{N,0} + \ket{N,0}\!\bra{0,N}$~\cite{Lee_2002_NOON, mitchell2004superNOON} can be used to estimate $t$ and $\omega$ at noise-enhanced precisions that saturate the quantum Cram\'er-Rao bound, with expressions for the corresponding quantum Fisher information in Eqs.~\eqref{eq:CatTime} and~\eqref{eq:CatFrequency}.

\vspace{12pt}
\noindent \textbf{Discussion}
\vspace{2pt}

The main results in this letter, Eqs.~\eqref{eq:NoiseAdvantage} and~\eqref{eq:NoiseAdvantageFreq}, identify regimes where incoherent dynamics enhance the metrological power of quantum sensors that estimate time or global parameters in a Hamiltonian. Tailored noise can increase the precision of an otherwise isolated quantum sensor. For example, for time estimation, designing noise-enhanced clocks would require 
    incoherent dynamics, as in Eq.~\eqref{eq:BoundFtime}, that influence the sensor over a time window $t - t_0 < \sqrt{2 \ln(2)}/\delta E$. A linearly ramped dephasing rate yields a precision improvement of $\dot \gamma (\delta L)^2/(\delta E)^2$. Other protocols could be tailored to the incoherent dynamics implementable in experimental platforms.

I have not studied the interplay between the suggested tailored noise and other potential noise sources that affect sensors in realistic conditions. Moreover, all derivations assume Lindblad operators that commute with the sensor's Hamiltonian. It would be interesting to explore whether non-commuting noise could sometimes serve as a metrological resource, and whether noise can enhance sensing of parameters beyond time and global fields.





\vspace{12pt}
\noindent \textbf{Acknowledgements} 
\vspace{2pt}

I thank Jacob Bringewatt for discussions about quantum metrology. This work was supported by the Laboratory Directed Research and Development program of Los Alamos National Laboratory under project number 20250902ER.





\bibliography{referencesPE}

\begin{thebibliography}{60}%
\makeatletter
\providecommand \@ifxundefined [1]{%
 \@ifx{#1\undefined}
}%
\providecommand \@ifnum [1]{%
 \ifnum #1\expandafter \@firstoftwo
 \else \expandafter \@secondoftwo
 \fi
}%
\providecommand \@ifx [1]{%
 \ifx #1\expandafter \@firstoftwo
 \else \expandafter \@secondoftwo
 \fi
}%
\providecommand \natexlab [1]{#1}%
\providecommand \enquote  [1]{``#1''}%
\providecommand \bibnamefont  [1]{#1}%
\providecommand \bibfnamefont [1]{#1}%
\providecommand \citenamefont [1]{#1}%
\providecommand \href@noop [0]{\@secondoftwo}%
\providecommand \href [0]{\begingroup \@sanitize@url \@href}%
\providecommand \@href[1]{\@@startlink{#1}\@@href}%
\providecommand \@@href[1]{\endgroup#1\@@endlink}%
\providecommand \@sanitize@url [0]{\catcode `\\12\catcode `\$12\catcode `\&12\catcode `\#12\catcode `\^12\catcode `\_12\catcode `\%12\relax}%
\providecommand \@@startlink[1]{}%
\providecommand \@@endlink[0]{}%
\providecommand \url  [0]{\begingroup\@sanitize@url \@url }%
\providecommand \@url [1]{\endgroup\@href {#1}{\urlprefix }}%
\providecommand \urlprefix  [0]{URL }%
\providecommand \Eprint [0]{\href }%
\providecommand \doibase [0]{https://doi.org/}%
\providecommand \selectlanguage [0]{\@gobble}%
\providecommand \bibinfo  [0]{\@secondoftwo}%
\providecommand \bibfield  [0]{\@secondoftwo}%
\providecommand \translation [1]{[#1]}%
\providecommand \BibitemOpen [0]{}%
\providecommand \bibitemStop [0]{}%
\providecommand \bibitemNoStop [0]{.\EOS\space}%
\providecommand \EOS [0]{\spacefactor3000\relax}%
\providecommand \BibitemShut  [1]{\csname bibitem#1\endcsname}%
\let\auto@bib@innerbib\@empty
\bibitem [{\citenamefont {Polzik}\ \emph {et~al.}(1992)\citenamefont {Polzik}, \citenamefont {Carri},\ and\ \citenamefont {Kimble}}]{PhysRevLett.68.3020}%
  \BibitemOpen
  \bibfield  {author} {\bibinfo {author} {\bibfnamefont {E.~S.}\ \bibnamefont {Polzik}}, \bibinfo {author} {\bibfnamefont {J.}~\bibnamefont {Carri}},\ and\ \bibinfo {author} {\bibfnamefont {H.~J.}\ \bibnamefont {Kimble}},\ }\bibfield  {title} {\bibinfo {title} {Spectroscopy with squeezed light},\ }\href {https://doi.org/10.1103/PhysRevLett.68.3020} {\bibfield  {journal} {\bibinfo  {journal} {Phys. Rev. Lett.}\ }\textbf {\bibinfo {volume} {68}},\ \bibinfo {pages} {3020} (\bibinfo {year} {1992})}\BibitemShut {NoStop}%
\bibitem [{\citenamefont {Collaboration}(2023)}]{LIGO}%
  \BibitemOpen
  \bibfield  {author} {\bibinfo {author} {\bibfnamefont {L.~O.~D.}\ \bibnamefont {Collaboration}} (\bibinfo {collaboration} {LIGO O4 Detector Collaboration}),\ }\bibfield  {title} {\bibinfo {title} {Broadband quantum enhancement of the {LIGO} detectors with frequency-dependent squeezing},\ }\href {https://doi.org/10.1103/PhysRevX.13.041021} {\bibfield  {journal} {\bibinfo  {journal} {Phys. Rev. X}\ }\textbf {\bibinfo {volume} {13}},\ \bibinfo {pages} {041021} (\bibinfo {year} {2023})}\BibitemShut {NoStop}%
\bibitem [{\citenamefont {Bollinger}\ \emph {et~al.}(1996)\citenamefont {Bollinger}, \citenamefont {Itano}, \citenamefont {Wineland},\ and\ \citenamefont {Heinzen}}]{Bollinger1996}%
  \BibitemOpen
  \bibfield  {author} {\bibinfo {author} {\bibfnamefont {J.~J.}\ \bibnamefont {Bollinger}}, \bibinfo {author} {\bibfnamefont {W.~M.}\ \bibnamefont {Itano}}, \bibinfo {author} {\bibfnamefont {D.~J.}\ \bibnamefont {Wineland}},\ and\ \bibinfo {author} {\bibfnamefont {D.~J.}\ \bibnamefont {Heinzen}},\ }\bibfield  {title} {\bibinfo {title} {Optimal frequency measurements with maximally correlated states},\ }\href {https://doi.org/10.1103/PhysRevA.54.R4649} {\bibfield  {journal} {\bibinfo  {journal} {Phys. Rev. A}\ }\textbf {\bibinfo {volume} {54}},\ \bibinfo {pages} {R4649} (\bibinfo {year} {1996})}\BibitemShut {NoStop}%
\bibitem [{\citenamefont {Giovannetti}\ \emph {et~al.}(2006)\citenamefont {Giovannetti}, \citenamefont {Lloyd},\ and\ \citenamefont {Maccone}}]{GiovannettiPRL2006}%
  \BibitemOpen
  \bibfield  {author} {\bibinfo {author} {\bibfnamefont {V.}~\bibnamefont {Giovannetti}}, \bibinfo {author} {\bibfnamefont {S.}~\bibnamefont {Lloyd}},\ and\ \bibinfo {author} {\bibfnamefont {L.}~\bibnamefont {Maccone}},\ }\bibfield  {title} {\bibinfo {title} {Quantum metrology},\ }\href {https://doi.org/10.1103/PhysRevLett.96.010401} {\bibfield  {journal} {\bibinfo  {journal} {Phys. Rev. Lett.}\ }\textbf {\bibinfo {volume} {96}},\ \bibinfo {pages} {010401} (\bibinfo {year} {2006})}\BibitemShut {NoStop}%
\bibitem [{\citenamefont {Kwon}\ \emph {et~al.}(2019)\citenamefont {Kwon}, \citenamefont {Tan}, \citenamefont {Volkoff},\ and\ \citenamefont {Jeong}}]{tan2019quantum}%
  \BibitemOpen
  \bibfield  {author} {\bibinfo {author} {\bibfnamefont {H.}~\bibnamefont {Kwon}}, \bibinfo {author} {\bibfnamefont {K.~C.}\ \bibnamefont {Tan}}, \bibinfo {author} {\bibfnamefont {T.}~\bibnamefont {Volkoff}},\ and\ \bibinfo {author} {\bibfnamefont {H.}~\bibnamefont {Jeong}},\ }\bibfield  {title} {\bibinfo {title} {Nonclassicality as a quantifiable resource for quantum metrology},\ }\href {https://doi.org/10.1103/PhysRevLett.122.040503} {\bibfield  {journal} {\bibinfo  {journal} {Phys. Rev. Lett.}\ }\textbf {\bibinfo {volume} {122}},\ \bibinfo {pages} {040503} (\bibinfo {year} {2019})}\BibitemShut {NoStop}%
\bibitem [{\citenamefont {Lostaglio}(2020)}]{LostaglioPRL2020}%
  \BibitemOpen
  \bibfield  {author} {\bibinfo {author} {\bibfnamefont {M.}~\bibnamefont {Lostaglio}},\ }\bibfield  {title} {\bibinfo {title} {Certifying quantum signatures in thermodynamics and metrology via contextuality of quantum linear response},\ }\href {https://doi.org/10.1103/PhysRevLett.125.230603} {\bibfield  {journal} {\bibinfo  {journal} {Phys. Rev. Lett.}\ }\textbf {\bibinfo {volume} {125}},\ \bibinfo {pages} {230603} (\bibinfo {year} {2020})}\BibitemShut {NoStop}%
\bibitem [{\citenamefont {Ehrenberg}\ \emph {et~al.}(2023)\citenamefont {Ehrenberg}, \citenamefont {Bringewatt},\ and\ \citenamefont {Gorshkov}}]{PhysRevResearch.5.033228}%
  \BibitemOpen
  \bibfield  {author} {\bibinfo {author} {\bibfnamefont {A.}~\bibnamefont {Ehrenberg}}, \bibinfo {author} {\bibfnamefont {J.}~\bibnamefont {Bringewatt}},\ and\ \bibinfo {author} {\bibfnamefont {A.~V.}\ \bibnamefont {Gorshkov}},\ }\bibfield  {title} {\bibinfo {title} {Minimum-entanglement protocols for function estimation},\ }\href {https://doi.org/10.1103/PhysRevResearch.5.033228} {\bibfield  {journal} {\bibinfo  {journal} {Phys. Rev. Res.}\ }\textbf {\bibinfo {volume} {5}},\ \bibinfo {pages} {033228} (\bibinfo {year} {2023})}\BibitemShut {NoStop}%
\bibitem [{\citenamefont {Brady}\ \emph {et~al.}(2024)\citenamefont {Brady}, \citenamefont {Wang}, \citenamefont {Albert}, \citenamefont {Gorshkov},\ and\ \citenamefont {Zhuang}}]{brady2024correlated}%
  \BibitemOpen
  \bibfield  {author} {\bibinfo {author} {\bibfnamefont {A.~J.}\ \bibnamefont {Brady}}, \bibinfo {author} {\bibfnamefont {Y.-X.}\ \bibnamefont {Wang}}, \bibinfo {author} {\bibfnamefont {V.~V.}\ \bibnamefont {Albert}}, \bibinfo {author} {\bibfnamefont {A.~V.}\ \bibnamefont {Gorshkov}},\ and\ \bibinfo {author} {\bibfnamefont {Q.}~\bibnamefont {Zhuang}},\ }\bibfield  {title} {\bibinfo {title} {Correlated noise estimation with quantum sensor networks},\ }\bibfield  {journal} {\bibinfo  {journal} {arXiv preprint arXiv:2412.17903}\ }\href {https://doi.org/10.48550/arXiv.2412.17903} {10.48550/arXiv.2412.17903} (\bibinfo {year} {2024})\BibitemShut {NoStop}%
\bibitem [{\citenamefont {Schlosshauer}(2005)}]{Schlosshauer}%
  \BibitemOpen
  \bibfield  {author} {\bibinfo {author} {\bibfnamefont {M.}~\bibnamefont {Schlosshauer}},\ }\bibfield  {title} {\bibinfo {title} {Decoherence, the measurement problem, and interpretations of quantum mechanics},\ }\href {https://doi.org/10.1103/RevModPhys.76.1267} {\bibfield  {journal} {\bibinfo  {journal} {Rev. Mod. Phys.}\ }\textbf {\bibinfo {volume} {76}},\ \bibinfo {pages} {1267} (\bibinfo {year} {2005})}\BibitemShut {NoStop}%
\bibitem [{\citenamefont {Haase}\ \emph {et~al.}(2016)\citenamefont {Haase}, \citenamefont {Smirne}, \citenamefont {Huelga}, \citenamefont {Kołodynski},\ and\ \citenamefont {Demkowicz-Dobrzanski}}]{precisionlimitsopen}%
  \BibitemOpen
  \bibfield  {author} {\bibinfo {author} {\bibfnamefont {J.~F.}\ \bibnamefont {Haase}}, \bibinfo {author} {\bibfnamefont {A.}~\bibnamefont {Smirne}}, \bibinfo {author} {\bibfnamefont {S.~F.}\ \bibnamefont {Huelga}}, \bibinfo {author} {\bibfnamefont {J.}~\bibnamefont {Kołodynski}},\ and\ \bibinfo {author} {\bibfnamefont {R.}~\bibnamefont {Demkowicz-Dobrzanski}},\ }\bibfield  {title} {\bibinfo {title} {Precision limits in quantum metrology with open quantum systems},\ }\href {https://doi.org/10.1515/qmetro-2018-0002} {\bibfield  {journal} {\bibinfo  {journal} {Quantum Meas. Quantum Metrol.}\ }\textbf {\bibinfo {volume} {5}},\ \bibinfo {pages} {13–39} (\bibinfo {year} {2016})}\BibitemShut {NoStop}%
\bibitem [{\citenamefont {Datta}(2025)}]{Datta2025}%
  \BibitemOpen
  \bibfield  {author} {\bibinfo {author} {\bibfnamefont {A.}~\bibnamefont {Datta}},\ }\bibfield  {title} {\bibinfo {title} {Sensing with quantum light: a perspective},\ }\bibfield  {journal} {\bibinfo  {journal} {Nanophotonics}\ }\href {https://doi.org/doi:10.1515/nanoph-2024-0649} {doi:10.1515/nanoph-2024-0649} (\bibinfo {year} {2025})\BibitemShut {NoStop}%
\bibitem [{\citenamefont {Escher}\ \emph {et~al.}(2011{\natexlab{a}})\citenamefont {Escher}, \citenamefont {de~Matos~Filho},\ and\ \citenamefont {Davidovich}}]{escher2011quantum}%
  \BibitemOpen
  \bibfield  {author} {\bibinfo {author} {\bibfnamefont {B.}~\bibnamefont {Escher}}, \bibinfo {author} {\bibfnamefont {R.}~\bibnamefont {de~Matos~Filho}},\ and\ \bibinfo {author} {\bibfnamefont {L.}~\bibnamefont {Davidovich}},\ }\bibfield  {title} {\bibinfo {title} {Quantum metrology for noisy systems},\ }\href {https://doi.org/10.1007/s13538-011-0037-y} {\bibfield  {journal} {\bibinfo  {journal} {Brazilian Journal of Physics}\ }\textbf {\bibinfo {volume} {41}},\ \bibinfo {pages} {229} (\bibinfo {year} {2011}{\natexlab{a}})}\BibitemShut {NoStop}%
\bibitem [{\citenamefont {Demkowicz-Dobrza{\'n}ski}\ \emph {et~al.}(2012)\citenamefont {Demkowicz-Dobrza{\'n}ski}, \citenamefont {Ko{\l}ody{\'n}ski},\ and\ \citenamefont {Gu{\c{t}}{\u{a}}}}]{demkowicz2012NatComm}%
  \BibitemOpen
  \bibfield  {author} {\bibinfo {author} {\bibfnamefont {R.}~\bibnamefont {Demkowicz-Dobrza{\'n}ski}}, \bibinfo {author} {\bibfnamefont {J.}~\bibnamefont {Ko{\l}ody{\'n}ski}},\ and\ \bibinfo {author} {\bibfnamefont {M.}~\bibnamefont {Gu{\c{t}}{\u{a}}}},\ }\bibfield  {title} {\bibinfo {title} {The elusive {Heisenberg} limit in quantum-enhanced metrology},\ }\href {https://doi.org/10.1038/ncomms2067} {\bibfield  {journal} {\bibinfo  {journal} {Nat. Commun.}\ }\textbf {\bibinfo {volume} {3}},\ \bibinfo {pages} {1063} (\bibinfo {year} {2012})}\BibitemShut {NoStop}%
\bibitem [{\citenamefont {Tsang}(2013)}]{tsang2013quantum}%
  \BibitemOpen
  \bibfield  {author} {\bibinfo {author} {\bibfnamefont {M.}~\bibnamefont {Tsang}},\ }\bibfield  {title} {\bibinfo {title} {Quantum metrology with open dynamical systems},\ }\href {https://doi.org/10.1088/1367-2630/15/7/073005} {\bibfield  {journal} {\bibinfo  {journal} {New J. Phys.}\ }\textbf {\bibinfo {volume} {15}},\ \bibinfo {pages} {073005} (\bibinfo {year} {2013})}\BibitemShut {NoStop}%
\bibitem [{\citenamefont {Demkowicz-Dobrza\ifmmode~\acute{n}\else \'{n}\fi{}ski}\ \emph {et~al.}(2017)\citenamefont {Demkowicz-Dobrza\ifmmode~\acute{n}\else \'{n}\fi{}ski}, \citenamefont {Czajkowski},\ and\ \citenamefont {Sekatski}}]{PhysRevX.7.041009}%
  \BibitemOpen
  \bibfield  {author} {\bibinfo {author} {\bibfnamefont {R.}~\bibnamefont {Demkowicz-Dobrza\ifmmode~\acute{n}\else \'{n}\fi{}ski}}, \bibinfo {author} {\bibfnamefont {J.}~\bibnamefont {Czajkowski}},\ and\ \bibinfo {author} {\bibfnamefont {P.}~\bibnamefont {Sekatski}},\ }\bibfield  {title} {\bibinfo {title} {Adaptive quantum metrology under general markovian noise},\ }\href {https://doi.org/10.1103/PhysRevX.7.041009} {\bibfield  {journal} {\bibinfo  {journal} {Phys. Rev. X}\ }\textbf {\bibinfo {volume} {7}},\ \bibinfo {pages} {041009} (\bibinfo {year} {2017})}\BibitemShut {NoStop}%
\bibitem [{\citenamefont {Escher}\ \emph {et~al.}(2011{\natexlab{b}})\citenamefont {Escher}, \citenamefont {de~Matos~Filho},\ and\ \citenamefont {Davidovich}}]{escher2011general}%
  \BibitemOpen
  \bibfield  {author} {\bibinfo {author} {\bibfnamefont {B.}~\bibnamefont {Escher}}, \bibinfo {author} {\bibfnamefont {R.~L.}\ \bibnamefont {de~Matos~Filho}},\ and\ \bibinfo {author} {\bibfnamefont {L.}~\bibnamefont {Davidovich}},\ }\bibfield  {title} {\bibinfo {title} {General framework for estimating the ultimate precision limit in noisy quantum-enhanced metrology},\ }\href {https://doi.org/doi.org/10.1038/nphys1958} {\bibfield  {journal} {\bibinfo  {journal} {Nat. Phys.}\ }\textbf {\bibinfo {volume} {7}},\ \bibinfo {pages} {406} (\bibinfo {year} {2011}{\natexlab{b}})}\BibitemShut {NoStop}%
\bibitem [{\citenamefont {Alipour}\ \emph {et~al.}(2014)\citenamefont {Alipour}, \citenamefont {Mehboudi},\ and\ \citenamefont {Rezakhani}}]{PhysRevLett.112.120405}%
  \BibitemOpen
  \bibfield  {author} {\bibinfo {author} {\bibfnamefont {S.}~\bibnamefont {Alipour}}, \bibinfo {author} {\bibfnamefont {M.}~\bibnamefont {Mehboudi}},\ and\ \bibinfo {author} {\bibfnamefont {A.~T.}\ \bibnamefont {Rezakhani}},\ }\bibfield  {title} {\bibinfo {title} {Quantum metrology in open systems: Dissipative cram\'er-rao bound},\ }\href {https://doi.org/10.1103/PhysRevLett.112.120405} {\bibfield  {journal} {\bibinfo  {journal} {Phys. Rev. Lett.}\ }\textbf {\bibinfo {volume} {112}},\ \bibinfo {pages} {120405} (\bibinfo {year} {2014})}\BibitemShut {NoStop}%
\bibitem [{\citenamefont {Zhou}\ and\ \citenamefont {Jiang}(2021)}]{Zhou}%
  \BibitemOpen
  \bibfield  {author} {\bibinfo {author} {\bibfnamefont {S.}~\bibnamefont {Zhou}}\ and\ \bibinfo {author} {\bibfnamefont {L.}~\bibnamefont {Jiang}},\ }\bibfield  {title} {\bibinfo {title} {Asymptotic theory of quantum channel estimation},\ }\href {https://doi.org/10.1103/PRXQuantum.2.010343} {\bibfield  {journal} {\bibinfo  {journal} {PRX Quantum}\ }\textbf {\bibinfo {volume} {2}},\ \bibinfo {pages} {010343} (\bibinfo {year} {2021})}\BibitemShut {NoStop}%
\bibitem [{\citenamefont {Len}\ \emph {et~al.}(2022)\citenamefont {Len}, \citenamefont {Gefen}, \citenamefont {Retzker},\ and\ \citenamefont {Ko{\l}ody{\'n}ski}}]{len2022quantum}%
  \BibitemOpen
  \bibfield  {author} {\bibinfo {author} {\bibfnamefont {Y.~L.}\ \bibnamefont {Len}}, \bibinfo {author} {\bibfnamefont {T.}~\bibnamefont {Gefen}}, \bibinfo {author} {\bibfnamefont {A.}~\bibnamefont {Retzker}},\ and\ \bibinfo {author} {\bibfnamefont {J.}~\bibnamefont {Ko{\l}ody{\'n}ski}},\ }\bibfield  {title} {\bibinfo {title} {Quantum metrology with imperfect measurements},\ }\href {https://doi.org/10.1038/s41467-022-33563-8} {\bibfield  {journal} {\bibinfo  {journal} {Nat. Commun.}\ }\textbf {\bibinfo {volume} {13}},\ \bibinfo {pages} {6971} (\bibinfo {year} {2022})}\BibitemShut {NoStop}%
\bibitem [{\citenamefont {Das}\ \emph {et~al.}(2025)\citenamefont {Das}, \citenamefont {G\'orecki},\ and\ \citenamefont {Demkowicz-Dobrza\ifmmode~\acute{n}\else \'{n}\fi{}ski}}]{PhysRevA.111.L020403}%
  \BibitemOpen
  \bibfield  {author} {\bibinfo {author} {\bibfnamefont {A.}~\bibnamefont {Das}}, \bibinfo {author} {\bibfnamefont {W.}~\bibnamefont {G\'orecki}},\ and\ \bibinfo {author} {\bibfnamefont {R.}~\bibnamefont {Demkowicz-Dobrza\ifmmode~\acute{n}\else \'{n}\fi{}ski}},\ }\bibfield  {title} {\bibinfo {title} {Universal time scalings of sensitivity in markovian quantum metrology},\ }\href {https://doi.org/10.1103/PhysRevA.111.L020403} {\bibfield  {journal} {\bibinfo  {journal} {Phys. Rev. A}\ }\textbf {\bibinfo {volume} {111}},\ \bibinfo {pages} {L020403} (\bibinfo {year} {2025})}\BibitemShut {NoStop}%
\bibitem [{\citenamefont {Matsuzaki}\ \emph {et~al.}(2011)\citenamefont {Matsuzaki}, \citenamefont {Benjamin},\ and\ \citenamefont {Fitzsimons}}]{PhysRevA.84.012103}%
  \BibitemOpen
  \bibfield  {author} {\bibinfo {author} {\bibfnamefont {Y.}~\bibnamefont {Matsuzaki}}, \bibinfo {author} {\bibfnamefont {S.~C.}\ \bibnamefont {Benjamin}},\ and\ \bibinfo {author} {\bibfnamefont {J.}~\bibnamefont {Fitzsimons}},\ }\bibfield  {title} {\bibinfo {title} {Magnetic field sensing beyond the standard quantum limit under the effect of decoherence},\ }\href {https://doi.org/10.1103/PhysRevA.84.012103} {\bibfield  {journal} {\bibinfo  {journal} {Phys. Rev. A}\ }\textbf {\bibinfo {volume} {84}},\ \bibinfo {pages} {012103} (\bibinfo {year} {2011})}\BibitemShut {NoStop}%
\bibitem [{\citenamefont {Chaves}\ \emph {et~al.}(2013)\citenamefont {Chaves}, \citenamefont {Brask}, \citenamefont {Markiewicz}, \citenamefont {Ko\l{}ody\ifmmode~\acute{n}\else \'{n}\fi{}ski},\ and\ \citenamefont {Ac\'{\i}n}}]{Noisymetro}%
  \BibitemOpen
  \bibfield  {author} {\bibinfo {author} {\bibfnamefont {R.}~\bibnamefont {Chaves}}, \bibinfo {author} {\bibfnamefont {J.~B.}\ \bibnamefont {Brask}}, \bibinfo {author} {\bibfnamefont {M.}~\bibnamefont {Markiewicz}}, \bibinfo {author} {\bibfnamefont {J.}~\bibnamefont {Ko\l{}ody\ifmmode~\acute{n}\else \'{n}\fi{}ski}},\ and\ \bibinfo {author} {\bibfnamefont {A.}~\bibnamefont {Ac\'{\i}n}},\ }\bibfield  {title} {\bibinfo {title} {Noisy metrology beyond the standard quantum limit},\ }\href {https://doi.org/10.1103/PhysRevLett.111.120401} {\bibfield  {journal} {\bibinfo  {journal} {Phys. Rev. Lett.}\ }\textbf {\bibinfo {volume} {111}},\ \bibinfo {pages} {120401} (\bibinfo {year} {2013})}\BibitemShut {NoStop}%
\bibitem [{\citenamefont {Koppenh\"ofer}\ \emph {et~al.}(2022)\citenamefont {Koppenh\"ofer}, \citenamefont {Groszkowski}, \citenamefont {Lau},\ and\ \citenamefont {Clerk}}]{PRXQuantum.3.030330}%
  \BibitemOpen
  \bibfield  {author} {\bibinfo {author} {\bibfnamefont {M.}~\bibnamefont {Koppenh\"ofer}}, \bibinfo {author} {\bibfnamefont {P.}~\bibnamefont {Groszkowski}}, \bibinfo {author} {\bibfnamefont {H.-K.}\ \bibnamefont {Lau}},\ and\ \bibinfo {author} {\bibfnamefont {A.}~\bibnamefont {Clerk}},\ }\bibfield  {title} {\bibinfo {title} {Dissipative superradiant spin amplifier for enhanced quantum sensing},\ }\href {https://doi.org/10.1103/PRXQuantum.3.030330} {\bibfield  {journal} {\bibinfo  {journal} {PRX Quantum}\ }\textbf {\bibinfo {volume} {3}},\ \bibinfo {pages} {030330} (\bibinfo {year} {2022})}\BibitemShut {NoStop}%
\bibitem [{\citenamefont {Niroula}\ \emph {et~al.}(2024)\citenamefont {Niroula}, \citenamefont {Dolde}, \citenamefont {Zheng}, \citenamefont {Bringewatt}, \citenamefont {Ehrenberg}, \citenamefont {Cox}, \citenamefont {Thompson}, \citenamefont {Gullans}, \citenamefont {Kolkowitz},\ and\ \citenamefont {Gorshkov}}]{PhysRevLett.133.080801}%
  \BibitemOpen
  \bibfield  {author} {\bibinfo {author} {\bibfnamefont {P.}~\bibnamefont {Niroula}}, \bibinfo {author} {\bibfnamefont {J.}~\bibnamefont {Dolde}}, \bibinfo {author} {\bibfnamefont {X.}~\bibnamefont {Zheng}}, \bibinfo {author} {\bibfnamefont {J.}~\bibnamefont {Bringewatt}}, \bibinfo {author} {\bibfnamefont {A.}~\bibnamefont {Ehrenberg}}, \bibinfo {author} {\bibfnamefont {K.~C.}\ \bibnamefont {Cox}}, \bibinfo {author} {\bibfnamefont {J.}~\bibnamefont {Thompson}}, \bibinfo {author} {\bibfnamefont {M.~J.}\ \bibnamefont {Gullans}}, \bibinfo {author} {\bibfnamefont {S.}~\bibnamefont {Kolkowitz}},\ and\ \bibinfo {author} {\bibfnamefont {A.~V.}\ \bibnamefont {Gorshkov}},\ }\bibfield  {title} {\bibinfo {title} {Quantum sensing with erasure qubits},\ }\href {https://doi.org/10.1103/PhysRevLett.133.080801} {\bibfield  {journal} {\bibinfo  {journal} {Phys. Rev. Lett.}\ }\textbf {\bibinfo {volume} {133}},\ \bibinfo {pages} {080801} (\bibinfo {year} {2024})}\BibitemShut {NoStop}%
\bibitem [{\citenamefont {D\"ur}\ \emph {et~al.}(2014)\citenamefont {D\"ur}, \citenamefont {Skotiniotis}, \citenamefont {Fr\"owis},\ and\ \citenamefont {Kraus}}]{QECPE1}%
  \BibitemOpen
  \bibfield  {author} {\bibinfo {author} {\bibfnamefont {W.}~\bibnamefont {D\"ur}}, \bibinfo {author} {\bibfnamefont {M.}~\bibnamefont {Skotiniotis}}, \bibinfo {author} {\bibfnamefont {F.}~\bibnamefont {Fr\"owis}},\ and\ \bibinfo {author} {\bibfnamefont {B.}~\bibnamefont {Kraus}},\ }\bibfield  {title} {\bibinfo {title} {Improved quantum metrology using quantum error correction},\ }\href {https://doi.org/10.1103/PhysRevLett.112.080801} {\bibfield  {journal} {\bibinfo  {journal} {Phys. Rev. Lett.}\ }\textbf {\bibinfo {volume} {112}},\ \bibinfo {pages} {080801} (\bibinfo {year} {2014})}\BibitemShut {NoStop}%
\bibitem [{\citenamefont {Chen}\ \emph {et~al.}(2024)\citenamefont {Chen}, \citenamefont {Chen}, \citenamefont {Liu}, \citenamefont {Miao},\ and\ \citenamefont {Yuan}}]{PhysRevLett.133.190801}%
  \BibitemOpen
  \bibfield  {author} {\bibinfo {author} {\bibfnamefont {H.}~\bibnamefont {Chen}}, \bibinfo {author} {\bibfnamefont {Y.}~\bibnamefont {Chen}}, \bibinfo {author} {\bibfnamefont {J.}~\bibnamefont {Liu}}, \bibinfo {author} {\bibfnamefont {Z.}~\bibnamefont {Miao}},\ and\ \bibinfo {author} {\bibfnamefont {H.}~\bibnamefont {Yuan}},\ }\bibfield  {title} {\bibinfo {title} {Quantum metrology enhanced by leveraging informative noise with error correction},\ }\href {https://doi.org/10.1103/PhysRevLett.133.190801} {\bibfield  {journal} {\bibinfo  {journal} {Phys. Rev. Lett.}\ }\textbf {\bibinfo {volume} {133}},\ \bibinfo {pages} {190801} (\bibinfo {year} {2024})}\BibitemShut {NoStop}%
\bibitem [{\citenamefont {Zhou}\ \emph {et~al.}(2018)\citenamefont {Zhou}, \citenamefont {Zhang}, \citenamefont {Preskill},\ and\ \citenamefont {Jiang}}]{QECPE2}%
  \BibitemOpen
  \bibfield  {author} {\bibinfo {author} {\bibfnamefont {S.}~\bibnamefont {Zhou}}, \bibinfo {author} {\bibfnamefont {M.}~\bibnamefont {Zhang}}, \bibinfo {author} {\bibfnamefont {J.}~\bibnamefont {Preskill}},\ and\ \bibinfo {author} {\bibfnamefont {L.}~\bibnamefont {Jiang}},\ }\bibfield  {title} {\bibinfo {title} {Achieving the {Heisenberg} limit in quantum metrology using quantum error correction},\ }\href {https://doi.org/10.1038/s41467-017-02510-3} {\bibfield  {journal} {\bibinfo  {journal} {Nat. Commun.}\ }\textbf {\bibinfo {volume} {9}},\ \bibinfo {pages} {78} (\bibinfo {year} {2018})}\BibitemShut {NoStop}%
\bibitem [{\citenamefont {Beau}\ and\ \citenamefont {del Campo}(2017)}]{adc}%
  \BibitemOpen
  \bibfield  {author} {\bibinfo {author} {\bibfnamefont {M.}~\bibnamefont {Beau}}\ and\ \bibinfo {author} {\bibfnamefont {A.}~\bibnamefont {del Campo}},\ }\bibfield  {title} {\bibinfo {title} {Nonlinear quantum metrology of many-body open systems},\ }\href {https://doi.org/10.1103/PhysRevLett.119.010403} {\bibfield  {journal} {\bibinfo  {journal} {Phys. Rev. Lett.}\ }\textbf {\bibinfo {volume} {119}},\ \bibinfo {pages} {010403} (\bibinfo {year} {2017})}\BibitemShut {NoStop}%
\bibitem [{\citenamefont {Beau}\ \emph {et~al.}(2017)\citenamefont {Beau}, \citenamefont {Chenu}, \citenamefont {Cao},\ and\ \citenamefont {del Campo}}]{Beau:17}%
  \BibitemOpen
  \bibfield  {author} {\bibinfo {author} {\bibfnamefont {M.}~\bibnamefont {Beau}}, \bibinfo {author} {\bibfnamefont {A.}~\bibnamefont {Chenu}}, \bibinfo {author} {\bibfnamefont {J.}~\bibnamefont {Cao}},\ and\ \bibinfo {author} {\bibfnamefont {A.}~\bibnamefont {del Campo}},\ }\bibfield  {title} {\bibinfo {title} {Quantum simulation and quantum metrology of many-body decoherence},\ }in\ \href {https://doi.org/10.1364/QIM.2017.QF5B.3} {\emph {\bibinfo {booktitle} {Quantum Information and Measurement (QIM) 2017}}}\ (\bibinfo  {publisher} {Optica Publishing Group},\ \bibinfo {year} {2017})\ p.\ \bibinfo {pages} {QF5B.3}\BibitemShut {NoStop}%
\bibitem [{\citenamefont {Yang}\ \emph {et~al.}(2022)\citenamefont {Yang}, \citenamefont {Pang}, \citenamefont {del Campo},\ and\ \citenamefont {Jordan}}]{PhysRevResearch.4.013133}%
  \BibitemOpen
  \bibfield  {author} {\bibinfo {author} {\bibfnamefont {J.}~\bibnamefont {Yang}}, \bibinfo {author} {\bibfnamefont {S.}~\bibnamefont {Pang}}, \bibinfo {author} {\bibfnamefont {A.}~\bibnamefont {del Campo}},\ and\ \bibinfo {author} {\bibfnamefont {A.~N.}\ \bibnamefont {Jordan}},\ }\bibfield  {title} {\bibinfo {title} {Super-heisenberg scaling in hamiltonian parameter estimation in the long-range kitaev chain},\ }\href {https://doi.org/10.1103/PhysRevResearch.4.013133} {\bibfield  {journal} {\bibinfo  {journal} {Phys. Rev. Res.}\ }\textbf {\bibinfo {volume} {4}},\ \bibinfo {pages} {013133} (\bibinfo {year} {2022})}\BibitemShut {NoStop}%
\bibitem [{\citenamefont {Deffner}(2025)}]{deffner}%
  \BibitemOpen
  \bibfield  {author} {\bibinfo {author} {\bibfnamefont {S.}~\bibnamefont {Deffner}},\ }\bibfield  {title} {\bibinfo {title} {Towards enhanced precision in thermometry with nonlinear qubits},\ }\href {https://doi.org/10.1088/2058-9565/adac05} {\bibfield  {journal} {\bibinfo  {journal} {QST}\ }\textbf {\bibinfo {volume} {10}},\ \bibinfo {pages} {025009} (\bibinfo {year} {2025})}\BibitemShut {NoStop}%
\bibitem [{\citenamefont {Zanardi}\ \emph {et~al.}(2008)\citenamefont {Zanardi}, \citenamefont {Paris},\ and\ \citenamefont {Campos~Venuti}}]{PhysRevA.78.042105}%
  \BibitemOpen
  \bibfield  {author} {\bibinfo {author} {\bibfnamefont {P.}~\bibnamefont {Zanardi}}, \bibinfo {author} {\bibfnamefont {M.~G.~A.}\ \bibnamefont {Paris}},\ and\ \bibinfo {author} {\bibfnamefont {L.}~\bibnamefont {Campos~Venuti}},\ }\bibfield  {title} {\bibinfo {title} {Quantum criticality as a resource for quantum estimation},\ }\href {https://doi.org/10.1103/PhysRevA.78.042105} {\bibfield  {journal} {\bibinfo  {journal} {Phys. Rev. A}\ }\textbf {\bibinfo {volume} {78}},\ \bibinfo {pages} {042105} (\bibinfo {year} {2008})}\BibitemShut {NoStop}%
\bibitem [{\citenamefont {Fr\'erot}\ and\ \citenamefont {Roscilde}(2018)}]{PhysRevLett.121.020402}%
  \BibitemOpen
  \bibfield  {author} {\bibinfo {author} {\bibfnamefont {I.}~\bibnamefont {Fr\'erot}}\ and\ \bibinfo {author} {\bibfnamefont {T.}~\bibnamefont {Roscilde}},\ }\bibfield  {title} {\bibinfo {title} {Quantum critical metrology},\ }\href {https://doi.org/10.1103/PhysRevLett.121.020402} {\bibfield  {journal} {\bibinfo  {journal} {Phys. Rev. Lett.}\ }\textbf {\bibinfo {volume} {121}},\ \bibinfo {pages} {020402} (\bibinfo {year} {2018})}\BibitemShut {NoStop}%
\bibitem [{\citenamefont {Ilias}\ \emph {et~al.}(2022)\citenamefont {Ilias}, \citenamefont {Yang}, \citenamefont {Huelga},\ and\ \citenamefont {Plenio}}]{PRXQuantum.3.010354}%
  \BibitemOpen
  \bibfield  {author} {\bibinfo {author} {\bibfnamefont {T.}~\bibnamefont {Ilias}}, \bibinfo {author} {\bibfnamefont {D.}~\bibnamefont {Yang}}, \bibinfo {author} {\bibfnamefont {S.~F.}\ \bibnamefont {Huelga}},\ and\ \bibinfo {author} {\bibfnamefont {M.~B.}\ \bibnamefont {Plenio}},\ }\bibfield  {title} {\bibinfo {title} {Criticality-enhanced quantum sensing via continuous measurement},\ }\href {https://doi.org/10.1103/PRXQuantum.3.010354} {\bibfield  {journal} {\bibinfo  {journal} {PRX Quantum}\ }\textbf {\bibinfo {volume} {3}},\ \bibinfo {pages} {010354} (\bibinfo {year} {2022})}\BibitemShut {NoStop}%
\bibitem [{\citenamefont {Yu}\ \emph {et~al.}(2024)\citenamefont {Yu}, \citenamefont {Nguyen},\ and\ \citenamefont {Nimmrichter}}]{yu2024criticality}%
  \BibitemOpen
  \bibfield  {author} {\bibinfo {author} {\bibfnamefont {M.}~\bibnamefont {Yu}}, \bibinfo {author} {\bibfnamefont {H.~C.}\ \bibnamefont {Nguyen}},\ and\ \bibinfo {author} {\bibfnamefont {S.}~\bibnamefont {Nimmrichter}},\ }\bibfield  {title} {\bibinfo {title} {Criticality-enhanced precision in phase thermometry},\ }\href {https://doi.org/10.1103/PhysRevResearch.6.043094} {\bibfield  {journal} {\bibinfo  {journal} {Phys. Rev. Res.}\ }\textbf {\bibinfo {volume} {6}},\ \bibinfo {pages} {043094} (\bibinfo {year} {2024})}\BibitemShut {NoStop}%
\bibitem [{\citenamefont {Ostermann}\ and\ \citenamefont {Gietka}(2024)}]{PhysRevA.109.L050601}%
  \BibitemOpen
  \bibfield  {author} {\bibinfo {author} {\bibfnamefont {L.}~\bibnamefont {Ostermann}}\ and\ \bibinfo {author} {\bibfnamefont {K.}~\bibnamefont {Gietka}},\ }\bibfield  {title} {\bibinfo {title} {Temperature-enhanced critical quantum metrology},\ }\href {https://doi.org/10.1103/PhysRevA.109.L050601} {\bibfield  {journal} {\bibinfo  {journal} {Phys. Rev. A}\ }\textbf {\bibinfo {volume} {109}},\ \bibinfo {pages} {L050601} (\bibinfo {year} {2024})}\BibitemShut {NoStop}%
\bibitem [{\citenamefont {Wang}\ \emph {et~al.}(2024)\citenamefont {Wang}, \citenamefont {Bringewatt}, \citenamefont {Seif}, \citenamefont {Brady}, \citenamefont {Oh},\ and\ \citenamefont {Gorshkov}}]{wang2024exponential}%
  \BibitemOpen
  \bibfield  {author} {\bibinfo {author} {\bibfnamefont {Y.-X.}\ \bibnamefont {Wang}}, \bibinfo {author} {\bibfnamefont {J.}~\bibnamefont {Bringewatt}}, \bibinfo {author} {\bibfnamefont {A.}~\bibnamefont {Seif}}, \bibinfo {author} {\bibfnamefont {A.~J.}\ \bibnamefont {Brady}}, \bibinfo {author} {\bibfnamefont {C.}~\bibnamefont {Oh}},\ and\ \bibinfo {author} {\bibfnamefont {A.~V.}\ \bibnamefont {Gorshkov}},\ }\bibfield  {title} {\bibinfo {title} {Exponential entanglement advantage in sensing correlated noise},\ }\bibfield  {journal} {\bibinfo  {journal} {arXiv preprint arXiv:2410.05878}\ }\href {https://doi.org/10.48550/arXiv.2410.05878} {10.48550/arXiv.2410.05878} (\bibinfo {year} {2024})\BibitemShut {NoStop}%
\bibitem [{\citenamefont {Peng}\ \emph {et~al.}(2024)\citenamefont {Peng}, \citenamefont {Zhu}, \citenamefont {Zhang},\ and\ \citenamefont {Zhang}}]{PhysRevLett.133.090801}%
  \BibitemOpen
  \bibfield  {author} {\bibinfo {author} {\bibfnamefont {J.-X.}\ \bibnamefont {Peng}}, \bibinfo {author} {\bibfnamefont {B.}~\bibnamefont {Zhu}}, \bibinfo {author} {\bibfnamefont {W.}~\bibnamefont {Zhang}},\ and\ \bibinfo {author} {\bibfnamefont {K.}~\bibnamefont {Zhang}},\ }\bibfield  {title} {\bibinfo {title} {Enhanced quantum metrology with non-phase-covariant noise},\ }\href {https://doi.org/10.1103/PhysRevLett.133.090801} {\bibfield  {journal} {\bibinfo  {journal} {Phys. Rev. Lett.}\ }\textbf {\bibinfo {volume} {133}},\ \bibinfo {pages} {090801} (\bibinfo {year} {2024})}\BibitemShut {NoStop}%
\bibitem [{\citenamefont {Verstraete}\ \emph {et~al.}(2009)\citenamefont {Verstraete}, \citenamefont {Wolf},\ and\ \citenamefont {Ignacio~Cirac}}]{verstraete2009quantum}%
  \BibitemOpen
  \bibfield  {author} {\bibinfo {author} {\bibfnamefont {F.}~\bibnamefont {Verstraete}}, \bibinfo {author} {\bibfnamefont {M.~M.}\ \bibnamefont {Wolf}},\ and\ \bibinfo {author} {\bibfnamefont {J.}~\bibnamefont {Ignacio~Cirac}},\ }\bibfield  {title} {\bibinfo {title} {Quantum computation and quantum-state engineering driven by dissipation},\ }\href {https://doi.org/10.1038/nphys1342} {\bibfield  {journal} {\bibinfo  {journal} {Nat. Phys.}\ }\textbf {\bibinfo {volume} {5}},\ \bibinfo {pages} {633} (\bibinfo {year} {2009})}\BibitemShut {NoStop}%
\bibitem [{\citenamefont {Pastawski}\ \emph {et~al.}(2011)\citenamefont {Pastawski}, \citenamefont {Clemente},\ and\ \citenamefont {Cirac}}]{PhysRevA.83.012304}%
  \BibitemOpen
  \bibfield  {author} {\bibinfo {author} {\bibfnamefont {F.}~\bibnamefont {Pastawski}}, \bibinfo {author} {\bibfnamefont {L.}~\bibnamefont {Clemente}},\ and\ \bibinfo {author} {\bibfnamefont {J.~I.}\ \bibnamefont {Cirac}},\ }\bibfield  {title} {\bibinfo {title} {Quantum memories based on engineered dissipation},\ }\href {https://doi.org/10.1103/PhysRevA.83.012304} {\bibfield  {journal} {\bibinfo  {journal} {Phys. Rev. A}\ }\textbf {\bibinfo {volume} {83}},\ \bibinfo {pages} {012304} (\bibinfo {year} {2011})}\BibitemShut {NoStop}%
\bibitem [{\citenamefont {Chenu}\ \emph {et~al.}(2017)\citenamefont {Chenu}, \citenamefont {Beau}, \citenamefont {Cao},\ and\ \citenamefont {del Campo}}]{PhysRevLett.118.140403}%
  \BibitemOpen
  \bibfield  {author} {\bibinfo {author} {\bibfnamefont {A.}~\bibnamefont {Chenu}}, \bibinfo {author} {\bibfnamefont {M.}~\bibnamefont {Beau}}, \bibinfo {author} {\bibfnamefont {J.}~\bibnamefont {Cao}},\ and\ \bibinfo {author} {\bibfnamefont {A.}~\bibnamefont {del Campo}},\ }\bibfield  {title} {\bibinfo {title} {Quantum simulation of generic many-body open system dynamics using classical noise},\ }\href {https://doi.org/10.1103/PhysRevLett.118.140403} {\bibfield  {journal} {\bibinfo  {journal} {Phys. Rev. Lett.}\ }\textbf {\bibinfo {volume} {118}},\ \bibinfo {pages} {140403} (\bibinfo {year} {2017})}\BibitemShut {NoStop}%
\bibitem [{\citenamefont {Harrington}\ \emph {et~al.}(2022)\citenamefont {Harrington}, \citenamefont {Mueller},\ and\ \citenamefont {Murch}}]{Harrington_2022}%
  \BibitemOpen
  \bibfield  {author} {\bibinfo {author} {\bibfnamefont {P.~M.}\ \bibnamefont {Harrington}}, \bibinfo {author} {\bibfnamefont {E.~J.}\ \bibnamefont {Mueller}},\ and\ \bibinfo {author} {\bibfnamefont {K.~W.}\ \bibnamefont {Murch}},\ }\bibfield  {title} {\bibinfo {title} {Engineered dissipation for quantum information science},\ }\href {https://doi.org/10.1038/s42254-022-00494-8} {\bibfield  {journal} {\bibinfo  {journal} {Nat. Rev. Phys.}\ }\textbf {\bibinfo {volume} {4}},\ \bibinfo {pages} {660–671} (\bibinfo {year} {2022})}\BibitemShut {NoStop}%
\bibitem [{\citenamefont {Sannia}\ \emph {et~al.}(2024)\citenamefont {Sannia}, \citenamefont {Tacchino}, \citenamefont {Tavernelli}, \citenamefont {Giorgi},\ and\ \citenamefont {Zambrini}}]{Sannia_2024}%
  \BibitemOpen
  \bibfield  {author} {\bibinfo {author} {\bibfnamefont {A.}~\bibnamefont {Sannia}}, \bibinfo {author} {\bibfnamefont {F.}~\bibnamefont {Tacchino}}, \bibinfo {author} {\bibfnamefont {I.}~\bibnamefont {Tavernelli}}, \bibinfo {author} {\bibfnamefont {G.~L.}\ \bibnamefont {Giorgi}},\ and\ \bibinfo {author} {\bibfnamefont {R.}~\bibnamefont {Zambrini}},\ }\bibfield  {title} {\bibinfo {title} {Engineered dissipation to mitigate barren plateaus},\ }\href {https://doi.org/10.1038/s41534-024-00875-0} {\bibfield  {journal} {\bibinfo  {journal} {npj Quantum Information}\ }\textbf {\bibinfo {volume} {10}},\ \bibinfo {pages} {81} (\bibinfo {year} {2024})}\BibitemShut {NoStop}%
\bibitem [{\citenamefont {Martinez-Azcona}\ \emph {et~al.}(2024)\citenamefont {Martinez-Azcona}, \citenamefont {Kundu}, \citenamefont {Saxena}, \citenamefont {del Campo},\ and\ \citenamefont {Chenu}}]{martinez2024quantum}%
  \BibitemOpen
  \bibfield  {author} {\bibinfo {author} {\bibfnamefont {P.}~\bibnamefont {Martinez-Azcona}}, \bibinfo {author} {\bibfnamefont {A.}~\bibnamefont {Kundu}}, \bibinfo {author} {\bibfnamefont {A.}~\bibnamefont {Saxena}}, \bibinfo {author} {\bibfnamefont {A.}~\bibnamefont {del Campo}},\ and\ \bibinfo {author} {\bibfnamefont {A.}~\bibnamefont {Chenu}},\ }\bibfield  {title} {\bibinfo {title} {Quantum dynamics with stochastic non-{Hermitian Hamiltonians}},\ }\href@noop {} {\bibfield  {journal} {\bibinfo  {journal} {arXiv}\ } (\bibinfo {year} {2024})},\ \Eprint {https://arxiv.org/abs/2407.07746} {2407.07746 [quant-ph]} \BibitemShut {NoStop}%
\bibitem [{\citenamefont {Helstrom}(1969)}]{helstrom1969quantum}%
  \BibitemOpen
  \bibfield  {author} {\bibinfo {author} {\bibfnamefont {C.~W.}\ \bibnamefont {Helstrom}},\ }\bibfield  {title} {\bibinfo {title} {Quantum detection and estimation theory},\ }\href {https://doi.org/10.1007/BF01007479} {\bibfield  {journal} {\bibinfo  {journal} {J. Stat. Phys.}\ }\textbf {\bibinfo {volume} {1}},\ \bibinfo {pages} {231} (\bibinfo {year} {1969})}\BibitemShut {NoStop}%
\bibitem [{\citenamefont {Liu}\ \emph {et~al.}(2019)\citenamefont {Liu}, \citenamefont {Yuan}, \citenamefont {Lu},\ and\ \citenamefont {Wang}}]{liu2019quantum}%
  \BibitemOpen
  \bibfield  {author} {\bibinfo {author} {\bibfnamefont {J.}~\bibnamefont {Liu}}, \bibinfo {author} {\bibfnamefont {H.}~\bibnamefont {Yuan}}, \bibinfo {author} {\bibfnamefont {X.-M.}\ \bibnamefont {Lu}},\ and\ \bibinfo {author} {\bibfnamefont {X.}~\bibnamefont {Wang}},\ }\bibfield  {title} {\bibinfo {title} {Quantum {Fisher} information matrix and multiparameter estimation},\ }\href {https://doi.org/10.1088/1751-8121/ab5d4d} {\bibfield  {journal} {\bibinfo  {journal} {J. Phys. A: Math. and Theor.}\ }\textbf {\bibinfo {volume} {53}},\ \bibinfo {pages} {023001} (\bibinfo {year} {2019})}\BibitemShut {NoStop}%
\bibitem [{\citenamefont {Paris}(2009)}]{paris2009quantum}%
  \BibitemOpen
  \bibfield  {author} {\bibinfo {author} {\bibfnamefont {M.~G.}\ \bibnamefont {Paris}},\ }\bibfield  {title} {\bibinfo {title} {Quantum estimation for quantum technology},\ }\href {https://doi.org/10.1142/S0219749909004839} {\bibfield  {journal} {\bibinfo  {journal} {Int. J. Quantum Inf.}\ }\textbf {\bibinfo {volume} {7}},\ \bibinfo {pages} {125} (\bibinfo {year} {2009})}\BibitemShut {NoStop}%
\bibitem [{\citenamefont {Braunstein}\ and\ \citenamefont {Caves}(1994)}]{BraunsteinCaves1994}%
  \BibitemOpen
  \bibfield  {author} {\bibinfo {author} {\bibfnamefont {S.~L.}\ \bibnamefont {Braunstein}}\ and\ \bibinfo {author} {\bibfnamefont {C.~M.}\ \bibnamefont {Caves}},\ }\bibfield  {title} {\bibinfo {title} {Statistical distance and the geometry of quantum states},\ }\href {https://doi.org/10.1103/PhysRevLett.72.3439} {\bibfield  {journal} {\bibinfo  {journal} {Phys. Rev. Lett.}\ }\textbf {\bibinfo {volume} {72}},\ \bibinfo {pages} {3439} (\bibinfo {year} {1994})}\BibitemShut {NoStop}%
\bibitem [{\citenamefont {Lee}\ \emph {et~al.}(2002{\natexlab{a}})\citenamefont {Lee}, \citenamefont {Kok}, \citenamefont {Cerf},\ and\ \citenamefont {Dowling}}]{NOON1}%
  \BibitemOpen
  \bibfield  {author} {\bibinfo {author} {\bibfnamefont {H.}~\bibnamefont {Lee}}, \bibinfo {author} {\bibfnamefont {P.}~\bibnamefont {Kok}}, \bibinfo {author} {\bibfnamefont {N.~J.}\ \bibnamefont {Cerf}},\ and\ \bibinfo {author} {\bibfnamefont {J.~P.}\ \bibnamefont {Dowling}},\ }\bibfield  {title} {\bibinfo {title} {Linear optics and projective measurements alone suffice to create large-photon-number path entanglement},\ }\href {https://doi.org/10.1103/PhysRevA.65.030101} {\bibfield  {journal} {\bibinfo  {journal} {Phys. Rev. A}\ }\textbf {\bibinfo {volume} {65}},\ \bibinfo {pages} {030101} (\bibinfo {year} {2002}{\natexlab{a}})}\BibitemShut {NoStop}%
\bibitem [{\citenamefont {Kok}\ \emph {et~al.}(2002)\citenamefont {Kok}, \citenamefont {Lee},\ and\ \citenamefont {Dowling}}]{NOON2}%
  \BibitemOpen
  \bibfield  {author} {\bibinfo {author} {\bibfnamefont {P.}~\bibnamefont {Kok}}, \bibinfo {author} {\bibfnamefont {H.}~\bibnamefont {Lee}},\ and\ \bibinfo {author} {\bibfnamefont {J.~P.}\ \bibnamefont {Dowling}},\ }\bibfield  {title} {\bibinfo {title} {Creation of large-photon-number path entanglement conditioned on photodetection},\ }\href {https://doi.org/10.1103/PhysRevA.65.052104} {\bibfield  {journal} {\bibinfo  {journal} {Phys. Rev. A}\ }\textbf {\bibinfo {volume} {65}},\ \bibinfo {pages} {052104} (\bibinfo {year} {2002})}\BibitemShut {NoStop}%
\bibitem [{\citenamefont {Salvatori}\ \emph {et~al.}(2014)\citenamefont {Salvatori}, \citenamefont {Mandarino},\ and\ \citenamefont {Paris}}]{PhysRevA.90.022111}%
  \BibitemOpen
  \bibfield  {author} {\bibinfo {author} {\bibfnamefont {G.}~\bibnamefont {Salvatori}}, \bibinfo {author} {\bibfnamefont {A.}~\bibnamefont {Mandarino}},\ and\ \bibinfo {author} {\bibfnamefont {M.~G.~A.}\ \bibnamefont {Paris}},\ }\bibfield  {title} {\bibinfo {title} {Quantum metrology in {L}ipkin-{M}eshkov-{G}lick critical systems},\ }\href {https://doi.org/10.1103/PhysRevA.90.022111} {\bibfield  {journal} {\bibinfo  {journal} {Phys. Rev. A}\ }\textbf {\bibinfo {volume} {90}},\ \bibinfo {pages} {022111} (\bibinfo {year} {2014})}\BibitemShut {NoStop}%
\bibitem [{\citenamefont {Erker}\ \emph {et~al.}(2017)\citenamefont {Erker}, \citenamefont {Mitchison}, \citenamefont {Silva}, \citenamefont {Woods}, \citenamefont {Brunner},\ and\ \citenamefont {Huber}}]{PhysRevX.7.031022}%
  \BibitemOpen
  \bibfield  {author} {\bibinfo {author} {\bibfnamefont {P.}~\bibnamefont {Erker}}, \bibinfo {author} {\bibfnamefont {M.~T.}\ \bibnamefont {Mitchison}}, \bibinfo {author} {\bibfnamefont {R.}~\bibnamefont {Silva}}, \bibinfo {author} {\bibfnamefont {M.~P.}\ \bibnamefont {Woods}}, \bibinfo {author} {\bibfnamefont {N.}~\bibnamefont {Brunner}},\ and\ \bibinfo {author} {\bibfnamefont {M.}~\bibnamefont {Huber}},\ }\bibfield  {title} {\bibinfo {title} {Autonomous quantum clocks: Does thermodynamics limit our ability to measure time?},\ }\href {https://doi.org/10.1103/PhysRevX.7.031022} {\bibfield  {journal} {\bibinfo  {journal} {Phys. Rev. X}\ }\textbf {\bibinfo {volume} {7}},\ \bibinfo {pages} {031022} (\bibinfo {year} {2017})}\BibitemShut {NoStop}%
\bibitem [{\citenamefont {Meier}\ \emph {et~al.}(2023)\citenamefont {Meier}, \citenamefont {Schwarzhans}, \citenamefont {Erker},\ and\ \citenamefont {Huber}}]{PhysRevLett.131.220201}%
  \BibitemOpen
  \bibfield  {author} {\bibinfo {author} {\bibfnamefont {F.}~\bibnamefont {Meier}}, \bibinfo {author} {\bibfnamefont {E.}~\bibnamefont {Schwarzhans}}, \bibinfo {author} {\bibfnamefont {P.}~\bibnamefont {Erker}},\ and\ \bibinfo {author} {\bibfnamefont {M.}~\bibnamefont {Huber}},\ }\bibfield  {title} {\bibinfo {title} {Fundamental accuracy-resolution trade-off for timekeeping devices},\ }\href {https://doi.org/10.1103/PhysRevLett.131.220201} {\bibfield  {journal} {\bibinfo  {journal} {Phys. Rev. Lett.}\ }\textbf {\bibinfo {volume} {131}},\ \bibinfo {pages} {220201} (\bibinfo {year} {2023})}\BibitemShut {NoStop}%
\bibitem [{\citenamefont {Egusquiza}\ \emph {et~al.}(1999)\citenamefont {Egusquiza}, \citenamefont {Garay},\ and\ \citenamefont {Raya}}]{EgusquizaClocks1999}%
  \BibitemOpen
  \bibfield  {author} {\bibinfo {author} {\bibfnamefont {I.~L.}\ \bibnamefont {Egusquiza}}, \bibinfo {author} {\bibfnamefont {L.~J.}\ \bibnamefont {Garay}},\ and\ \bibinfo {author} {\bibfnamefont {J.~M.}\ \bibnamefont {Raya}},\ }\bibfield  {title} {\bibinfo {title} {Quantum evolution according to real clocks},\ }\href {https://doi.org/10.1103/PhysRevA.59.3236} {\bibfield  {journal} {\bibinfo  {journal} {Phys. Rev. A}\ }\textbf {\bibinfo {volume} {59}},\ \bibinfo {pages} {3236} (\bibinfo {year} {1999})}\BibitemShut {NoStop}%
\bibitem [{\citenamefont {Gambini}\ \emph {et~al.}(2004)\citenamefont {Gambini}, \citenamefont {Porto},\ and\ \citenamefont {Pullin}}]{GambiniClocks2004}%
  \BibitemOpen
  \bibfield  {author} {\bibinfo {author} {\bibfnamefont {R.}~\bibnamefont {Gambini}}, \bibinfo {author} {\bibfnamefont {R.~A.}\ \bibnamefont {Porto}},\ and\ \bibinfo {author} {\bibfnamefont {J.}~\bibnamefont {Pullin}},\ }\bibfield  {title} {\bibinfo {title} {Realistic clocks, universal decoherence, and the black hole information paradox},\ }\href {https://doi.org/10.1103/PhysRevLett.93.240401} {\bibfield  {journal} {\bibinfo  {journal} {Phys. Rev. Lett.}\ }\textbf {\bibinfo {volume} {93}},\ \bibinfo {pages} {240401} (\bibinfo {year} {2004})}\BibitemShut {NoStop}%
\bibitem [{\citenamefont {Xuereb}\ \emph {et~al.}(2023)\citenamefont {Xuereb}, \citenamefont {Erker}, \citenamefont {Meier}, \citenamefont {Mitchison},\ and\ \citenamefont {Huber}}]{HuberClocks2023}%
  \BibitemOpen
  \bibfield  {author} {\bibinfo {author} {\bibfnamefont {J.}~\bibnamefont {Xuereb}}, \bibinfo {author} {\bibfnamefont {P.}~\bibnamefont {Erker}}, \bibinfo {author} {\bibfnamefont {F.}~\bibnamefont {Meier}}, \bibinfo {author} {\bibfnamefont {M.~T.}\ \bibnamefont {Mitchison}},\ and\ \bibinfo {author} {\bibfnamefont {M.}~\bibnamefont {Huber}},\ }\bibfield  {title} {\bibinfo {title} {Impact of imperfect timekeeping on quantum control},\ }\href {https://doi.org/10.1103/PhysRevLett.131.160204} {\bibfield  {journal} {\bibinfo  {journal} {Phys. Rev. Lett.}\ }\textbf {\bibinfo {volume} {131}},\ \bibinfo {pages} {160204} (\bibinfo {year} {2023})}\BibitemShut {NoStop}%
\bibitem [{\citenamefont {G\'orecki}\ \emph {et~al.}(2020)\citenamefont {G\'orecki}, \citenamefont {Demkowicz-Dobrza\ifmmode~\acute{n}\else \'{n}\fi{}ski}, \citenamefont {Wiseman},\ and\ \citenamefont {Berry}}]{PhysRevLett.124.030501}%
  \BibitemOpen
  \bibfield  {author} {\bibinfo {author} {\bibfnamefont {W.}~\bibnamefont {G\'orecki}}, \bibinfo {author} {\bibfnamefont {R.}~\bibnamefont {Demkowicz-Dobrza\ifmmode~\acute{n}\else \'{n}\fi{}ski}}, \bibinfo {author} {\bibfnamefont {H.~M.}\ \bibnamefont {Wiseman}},\ and\ \bibinfo {author} {\bibfnamefont {D.~W.}\ \bibnamefont {Berry}},\ }\bibfield  {title} {\bibinfo {title} {$\ensuremath{\pi}$-corrected {Heisenberg} limit},\ }\href {https://doi.org/10.1103/PhysRevLett.124.030501} {\bibfield  {journal} {\bibinfo  {journal} {Phys. Rev. Lett.}\ }\textbf {\bibinfo {volume} {124}},\ \bibinfo {pages} {030501} (\bibinfo {year} {2020})}\BibitemShut {NoStop}%
\bibitem [{\citenamefont {Belliardo}\ and\ \citenamefont {Giovannetti}(2020)}]{PhysRevA.102.042613}%
  \BibitemOpen
  \bibfield  {author} {\bibinfo {author} {\bibfnamefont {F.}~\bibnamefont {Belliardo}}\ and\ \bibinfo {author} {\bibfnamefont {V.}~\bibnamefont {Giovannetti}},\ }\bibfield  {title} {\bibinfo {title} {Achieving {Heisenberg} scaling with maximally entangled states: An analytic upper bound for the attainable root-mean-square error},\ }\href {https://doi.org/10.1103/PhysRevA.102.042613} {\bibfield  {journal} {\bibinfo  {journal} {Phys. Rev. A}\ }\textbf {\bibinfo {volume} {102}},\ \bibinfo {pages} {042613} (\bibinfo {year} {2020})}\BibitemShut {NoStop}%
\bibitem [{\citenamefont {Lee}\ \emph {et~al.}(2002{\natexlab{b}})\citenamefont {Lee}, \citenamefont {Kok},\ and\ \citenamefont {Dowling}}]{Lee_2002_NOON}%
  \BibitemOpen
  \bibfield  {author} {\bibinfo {author} {\bibfnamefont {H.}~\bibnamefont {Lee}}, \bibinfo {author} {\bibfnamefont {P.}~\bibnamefont {Kok}},\ and\ \bibinfo {author} {\bibfnamefont {J.~P.}\ \bibnamefont {Dowling}},\ }\bibfield  {title} {\bibinfo {title} {A quantum rosetta stone for interferometry},\ }\href {https://doi.org/10.1080/0950034021000011536} {\bibfield  {journal} {\bibinfo  {journal} {J. Mod. Opt.}\ }\textbf {\bibinfo {volume} {49}},\ \bibinfo {pages} {2325–2338} (\bibinfo {year} {2002}{\natexlab{b}})}\BibitemShut {NoStop}%
\bibitem [{\citenamefont {Mitchell}\ \emph {et~al.}(2004)\citenamefont {Mitchell}, \citenamefont {Lundeen},\ and\ \citenamefont {Steinberg}}]{mitchell2004superNOON}%
  \BibitemOpen
  \bibfield  {author} {\bibinfo {author} {\bibfnamefont {M.~W.}\ \bibnamefont {Mitchell}}, \bibinfo {author} {\bibfnamefont {J.~S.}\ \bibnamefont {Lundeen}},\ and\ \bibinfo {author} {\bibfnamefont {A.~M.}\ \bibnamefont {Steinberg}},\ }\bibfield  {title} {\bibinfo {title} {Super-resolving phase measurements with a multiphoton entangled state},\ }\href {https://doi.org/10.1038/nature02493} {\bibfield  {journal} {\bibinfo  {journal} {Nature}\ }\textbf {\bibinfo {volume} {429}},\ \bibinfo {pages} {161} (\bibinfo {year} {2004})}\BibitemShut {NoStop}%
\end{thebibliography}%

\clearpage
\newpage

\onecolumngrid

\section*{Appendix}
\setcounter{secnumdepth}{1}
\renewcommand{\thesection}{A\arabic{section}}
\setcounter{equation}{0}
\renewcommand{\theequation}{A\arabic{equation}}


This Appendix includes useful expressions for the quantum Fisher information (Sec.~\ref{app:QFI}) and derivations of some of the results in the main text. Equations~\eqref{eq:BoundFtime} and~\eqref{eq:FtimeOpen} of the main text are proven in Sec.~\ref{app:QFItime}. Equations~\eqref{eq:BoundFfrequency} and~\eqref{eq:FfrequencyOpen}  of the main text are proven in Sec.~\ref{app:QFIfrequency}. Section~\ref{app:optimalestimators} details a protocol that saturates the quantum Cram\'er-Rao bound in noise-enhanced metrological settings.

\section{Background: the quantum Fisher information}
\label{app:QFI}

For a system in a state $\rho$, the symmetric logarithmic derivative $\mathcal L(\lambda)$ about a parameter $\lambda$ is defined by  
\begin{align}
\label{eq-app:SLD}
    \frac{\partial \rho}{\partial \lambda} = \frac{1}{2}  \{ \mathcal L(\lambda),\rho\}. 
\end{align}
The quantum Fisher information about $\lambda$ is $F(\lambda) = \langle \mathcal L^2(\lambda) \rangle = \tr{\rho \mathcal L^2(\lambda)}$~\cite{liu2019quantum}. For any $\rho$, 
\begin{align}
    \tr{ \left( \frac{\partial \rho}{\partial \lambda} \right)^2} &= \frac{1}{4} \tr{ \{ \mathcal L(\lambda),\rho\} \{ \mathcal L(\lambda),\rho\}  } = \frac{1}{2} \tr{\rho^2 \mathcal L^2 (\lambda)}  + \frac{1}{2} \tr{\rho \mathcal L (\lambda) \rho \mathcal L (\lambda)} \nonumber \\
    &\leq \frac{1}{2} \tr{\rho^2 \mathcal L^2 (\lambda)}  + \frac{1}{2} \tr{\rho^2 \mathcal L^2 (\lambda) } \leq \tr{\rho \mathcal L^2 (\lambda)} = F(\lambda). 
\end{align}
I used the Cauchy-Schwarz inequality and the fact that the eigenvalues of $\rho$ are positive and normalized. Thus, 
\begin{align}
\label{eq-app:BoundF}
   F(\lambda) \geq \tr{ \left( \frac{\partial \rho}{\partial \lambda} \right)^2}. 
\end{align}

For pure states $\rho^2 = \rho$, a tighter expression relates the quantum Fisher information to the state's change with $\lambda$, $F(\lambda) = 2 \tr{ \left( \frac{\partial \rho}{\partial \lambda} \right)^2}$, which is proven by using Eq.~\eqref{eq-app:SLD} and the fact that $\langle \mathcal L(\lambda) \rangle = 0$ [which can, in turn, be seen by taking a trace on Eq.~\eqref{eq-app:SLD}].

In the eigenbasis of $\rho$, the quantum Fisher information about a parameter $\lambda$ can be written as
\begin{align}
\label{eq-app:qFisher}
F(\lambda) =  \sum_\alpha \frac{1}{p_\alpha} \left( \frac{\partial p_\alpha}{\partial \lambda} \right)^2  + 4 \sum_\alpha p_\alpha \left( \frac{\partial \bra{\alpha}}{\partial \lambda} \right) \left( \frac{\partial \ket{\alpha} }{  \partial \lambda } \right) - 8 \sum_{\alpha \beta} \frac{ p_\alpha p_\beta }{p_\alpha + p_\beta} \left| \bra{\beta} \frac{\partial \ket{\alpha}}{\partial \lambda}\right|^2, 
\end{align}
where $p_\alpha$ and $\ket{\alpha}$ denote the non-zero eigenvalues and the eigenvectors of $\rho$, respectively~\cite{liu2019quantum}.

\section{The quantum Fisher information about time under decoherent dynamics}
\label{app:QFItime}

Consider the time-dependent state $\rho_t$ that evolves unitarily under Hamiltonian $H$ until $t < t_0$. Moreover, assume that $\rho_{t\leq t_0}^2 = \rho_{t\leq t_0}$ is pure. Then, at $t < t_0$, the quantum Fisher information about time is $F_\cl (t) = 4 \var(H)$~\cite{BraunsteinCaves1994}.

Starting at time $t_0$, the state evolves following the Lindblad master equation 
\begin{align}
\label{eq-app:Lindblad}
\frac{d\rho_t}{dt} = -i [H,\rho_t] - \gamma_t [L,[L,\rho_t]],
\end{align}
which decoheres the state as it evolves. I assume Hermitian Lindblad operators $L$ such that $[L,H] = 0$ with a time-dependent dephasing rate $\gamma_t$.
Then, 
\begin{align}
\label{eq-app:Aux1}
    \tr{\left(\frac{\partial \rho_t}{\partial t}\right)^2} &= \tr{[H,\rho_t][\rho_t,H]} + 2i \gamma_t \tr{[H,\rho_t][L,[L,\rho_t]]} + \gamma_t^2 \tr{[L,[L,\rho_t]][L,[L,\rho_t]]}.
\end{align}
Since $[L,H] = 0$, the second term in Eq.~\eqref{eq-app:Aux1} is
\begin{align}
    \label{eq-app:Aux2}
    \tr{[H,\rho_t][L,[L,\rho_t]]} &= \tr{\big( H\rho_t - \rho_t H\big) \big( L^2 \rho_t + \rho_t L^2 - 2 L\rho_t L \big)}  = 0.
\end{align}
Then, Eq.~\eqref{eq-app:BoundF} implies that the Fisher information about the parameter $\lambda = t$ satisfies
\begin{align}
\label{eq-app:BoundFtimeaux}
    F_\open(t) \geq \left\| [H,\rho_t] \right\|_2^2 + \gamma_t^2 \left\| [L,[L,\rho_t]] \right\|_2^2,
\end{align}
where $\|A\|_2 \coloneqq \sqrt{\tr{A A^\dag} }$ is the Hilbert-Schmidt operator norm. This proves Eq.~\eqref{eq:BoundFtime} in the main text.

\subsection{The quantum Fisher information about time for a cat state}

Consider a sensor network initialized in a cat state 
\begin{align}
\label{eq-app:Cat}
\ket{\psi} = \frac{1}{\sqrt{2}} \big( \ket{E_0} + \ket{E_1} \big),
\end{align}
where $\ket{E_0}$ and $\ket{E_1}$ are eigenvectors of H y L. (Recall that $[H,L] = 0$ so there exists a joint eigenbasis of both operators.) The eigenvalue differences of $H$ and $L$ are $\delta E \coloneqq H \ket{E_1} - H \ket{E_0}$ and $\delta L \coloneqq L \ket{E_1} - L \ket{E_0}$, respectively.

Under the dynamics in Eq.~\eqref{eq-app:Lindblad}, the state at time $t$ is
\begin{align}
\label{eq-app:state}
\rho_t = \frac{1}{2} \Big[ \ket{E_0}\!\bra{E_0} + \ket{E_1}\!\bra{E_1} +  e^{i \delta E t} e^{- \delta L^2 \int_{t_0}^t \gamma_s ds } \ket{E_0}\!\bra{E_1} +  e^{-i  \delta E t} e^{- \delta L^2 \int_{t_0}^t \gamma_s ds} \ket{E_1}\!\bra{E_0}  \Big].
\end{align}

The eigenvectors of $\rho_t$ are 
\begin{subequations}
\label{eq-app:eigenstates}
    \begin{align}
\ket{\phi}  &= \frac{1}{\sqrt{2}} \left( e^{ i \delta E t/2} \ket{E_0} + e^{- i \delta E t/2} \ket{E_1} \right), \\
\ket{\phi_*} &= \frac{1}{\sqrt{2}} \left( e^{ i \delta E t/2} \ket{E_0} - e^{- i \delta E t/2} \ket{E_1} \right),
\end{align}
\end{subequations}
and the corresponding eigenvalues are
\begin{subequations}
\label{eq-app:probs}
    \begin{align}
p &= \bra{\phi} \rho_t \ket{\phi} = 1/4 + 1/4 + 1/4 e^{-  \delta L^2 \int_{t_0}^t \gamma_s ds } + 1/4 e^{- \delta L^2 \int_{t_0}^t \gamma_s ds } = \frac{1}{2} \left( 1 + e^{- \delta L \int_{t_0}^t \gamma_s ds} \right), \\
p_* &= \bra{\phi_*} \rho_t \ket{\phi_*} = 1/4 + 1/4 - 1/4 e^{-  \delta L^2 \int_{t_0}^t \gamma_s ds} - 1/4 e^{- \delta L^2 \int_{t_0}^t \gamma_s ds} = \frac{1}{2} \left( 1 - e^{- \delta L^2 \int_{t_0}^t \gamma_s ds } \right).
\end{align}
\end{subequations}
The time derivatives of the eigenvectors and eigenvalues are
\begin{subequations}
    \label{eq-app:timederiveigs}
    \begin{align}
        \frac{\partial }{\partial t} \ket{\phi}&=  i\frac{\delta E}{2} \frac{1}{\sqrt{2}} \left( e^{ i \delta E t/2} \ket{E_0} - e^{- i \delta E t/2} \ket{E_1} \right) = i\frac{\delta E}{2} \ket{\phi_*}, \\ 
        \frac{\partial }{\partial t} \ket{\phi_*} &=  i\frac{\delta E}{2} \frac{1}{\sqrt{2}} \left( e^{ i \delta E t/2} \ket{E_0} + e^{- i \delta E t/2} \ket{E_1} \right) = i\frac{\delta E}{2} \ket{\phi},  \\ 
        \frac{\partial }{\partial t} p &=  -\frac{1}{2} \gamma_t \delta L^2 e^{- \delta L^2 \int_{t_0}^t \gamma_s ds},  \\ 
        \frac{\partial }{\partial t} p_*&= \frac{1}{2} \gamma_t \delta L^2 e^{-  \delta L^2 \int_{t_0}^t \gamma_s ds}.   
    \end{align}
\end{subequations}

Using Eq.~\eqref{eq-app:qFisher} and Eqs.~\eqref{eq-app:eigenstates} through~\eqref{eq-app:timederiveigs}, the quantum Fisher information about time is
\begin{align}
\label{eq-app:FisherTimeCat}
   F_\open(t) &=    \sum_\alpha \frac{1}{p_\alpha} \left( \frac{\partial p_\alpha}{\partial t} \right)^2  + 4 \sum_\alpha p_\alpha \left( \frac{\partial \bra{\alpha}}{\partial t} \right) \left( \frac{\partial \ket{\alpha} }{  \partial t} \right) - 8 \sum_{\alpha \beta} \frac{ p_\alpha p_\beta }{p_\alpha + p_\beta} \left| \bra{\beta} \frac{\partial \ket{\alpha}}{\partial t}\right|^2 \nonumber \\
   &= \left( \frac{1}{2} \gamma_t \delta L^2 e^{-  \delta L^2 \int_{t_0}^t \gamma_s ds} \right)^2 \frac{1}{p p_*} + 4 \frac{\delta E^2}{4} - 16 p p_* \frac{\delta E^2}{4}  \nonumber \\ 
   &  =  \gamma_t^2 \delta L^4 e^{- 2 \delta L^2 \int_{t_0}^t \gamma_s ds}  \frac{1}{\left( 1 + e^{- \delta L^2 \int_{t_0}^t \gamma_s ds} \right) \left( 1 - e^{- \delta L^2 \int_{t_0}^t \gamma_s ds} \right)}  \nonumber \\
   & +   \delta E^2 -  \left( 1 + e^{- \delta L^2 \int_{t_0}^t \gamma_s ds} \right) \left( 1 - e^{- \delta L^2 \int_{t_0}^t \gamma_s ds} \right) \delta E^2   \nonumber \\
   & =  \gamma_t^2 \delta L^4 \frac{1}{e^{2 \delta L^2 \int_{t_0}^t \gamma_s ds}   - 1} + \delta E^2 - \delta E^2\left( 1 - e^{- 2 \delta L^2 \int_{t_0}^t \gamma_s ds} \right) \nonumber \\
   & =  \gamma_t^2 \delta L^4 \frac{1}{e^{2 \delta L^2 \int_{t_0}^t \gamma_s ds}   - 1} + \delta E^2  e^{- 2 \delta L^2 \int_{t_0}^t \gamma_s ds} \nonumber \\
   & =  \left( \gamma_t^2 \, \delta L^4 \frac{1}{1 - e^{-2 \, \delta L^2 \int_{t_0}^t \gamma_s ds} } + \delta E^2  \right) e^{- 2 \,  \delta L^2 \int_{t_0}^t \gamma_s ds}. 
\end{align}
This proves Eq.~\eqref{eq:FtimeOpen} in the main text.

\subsection{A regime where noise enhances precision of time estimation}

Let $\gamma_t = \dot \gamma (t-t_0)$, with constant $\dot \gamma$. Then 
\begin{align}
    \int_{t_0}^t \gamma_s ds = \dot \gamma \int_0^{t-t_0} s ds = \dot\gamma \frac{(t-t_0)^2}{2}.
\end{align}
Further, assume that 
\begin{align}
2(\delta L)^2 \int_{t_0}^t \gamma_s ds = (\delta L)^2 \dot \gamma (t-t_0)^2 = \ln(2),
\end{align}
which sets the exponential prefactor in Eq.~\eqref{eq-app:FisherTimeCat} to be $1/2$. Equation~\eqref{eq-app:FisherTimeCat} becomes
\begin{align}
    F_\open(t) = \left(     \frac{ \gamma_t^2 \, \delta L^4 }{ 1/2}  + \delta E^2   \right) \frac{1}{2}  =   \dot\gamma^2 (t-t_0)^2 (\delta L)^4   + \frac{1}{2}(\delta E)^2 = \dot \gamma (\delta L)^2 + \frac{1}{2}(\delta E)^2.
\end{align}
Thus, $F_\open(t)$ is larger than $F_\cl(t) = (\delta E)^2$ whenever the following conditions hold:
\begin{subequations}
    \begin{align}
        \dot \gamma (\delta L)^2 \geq \frac{(\delta E)^2}{2}, \qquad \qquad
        t-t_0 = \frac{\sqrt{\ln(2)}}{\sqrt{\dot \gamma} \delta L}.
    \end{align}
\end{subequations}

\section{The quantum Fisher information about a global Hamiltonian parameter under energy decoherence}
\label{app:QFIfrequency}

Let $\rho_t$ be a state that, starting at $t = t_0$, evolves following the Lindblad master equation 
\begin{align}
\label{eq-app:LindbladAux}
\frac{d\rho_t}{dt} = -i [H,\rho_t] - \gamma_t [H,[H,\rho_t]],  
\end{align}
with Lindblad operator $L = H$. Consider one wants to learn a global parameter $\omega$ in the Hamiltonian, $H = \omega h$, and that the eigenvalues of the Hermitian operator $h$ are $\epsilon_j$. The Hamiltonian's eigenvalues are $E_j = \omega \epsilon_j$, with corresponding eigenvectors $\ket{j}$.

The solution to Eq.~\eqref{eq-app:LindbladAux} is
\begin{align}
    \rho_t = \sum_{jk} \rho_{jk} e^{-i \omega (\epsilon_j - \epsilon_k) t} e^{- \omega^2 (\epsilon_j - \epsilon_k)^2 \int_{t_0}^t \gamma_s ds} \ket{j}\!\bra{k},
\end{align}
where $\rho_{jk} = \bra{j} \rho_0 \ket{k}$ are the initial state's matrix elements in the energy eigenbasis.

Then, 
\begin{align}
    \frac{\partial \rho_t}{\partial \omega} &= \sum_{jk} \rho_{jk}\big( -i (\epsilon_j -\epsilon_k) t\big) e^{-i \omega (\epsilon_j - \epsilon_k) t} e^{-\omega^2 (\epsilon_j - \epsilon_k)^2 \int_{t_0}^t \gamma_s ds} \ket{j}\!\bra{k} \nonumber \\
    &+ \sum_{jk} \rho_{jk} \left( -  2 \omega (\epsilon_j - \epsilon_k)^2 \int_{t_0}^t \gamma_s ds \right) e^{-i \omega (\epsilon_j - \epsilon_k) t} e^{- \omega^2 (\epsilon_j - \epsilon_k)^2 \int_{t_0}^t \gamma_s ds} \ket{j}\!\bra{k} \nonumber \\
    &= -i \frac{t}{\omega} [H,\rho_t] - \frac{2 \int_{t_0}^t \gamma_s ds }{\omega} [H,[H,\rho_t]].
\end{align}
A similar derivation to that of Eq.~\eqref{eq-app:BoundFtimeaux} thus shows that the quantum Fisher information about the parameter $\omega$ is
\begin{align}
\label{eq-app:BoundFfrequency}
    F_\open(\omega) \geq \frac{t^2}{\omega^2} \left\| [H,\rho_t]  \right\|_2^2  +  \frac{4 \left( \int_{t_0}^t \gamma_s ds\right)^2 }{\omega^2} \left\| [H,[H,\rho_t]] \right\|_2^2.
\end{align}
This proves Eq.~\eqref{eq:BoundFfrequency} in the main text.

\subsection{The quantum Fisher information about a global Hamiltonian parameter for a cat state}

If starting in a cat state as in Eq.~\eqref{eq-app:Cat} and suffering from energy decoherence, 
the state at time $t$ is
\begin{align}
\rho_t = \frac{1}{2} \Big[ \ket{E_0}\!\bra{E_0} + \ket{E_1}\!\bra{E_1} +  e^{i \omega \delta \epsilon t} e^{- \omega^2 \delta \epsilon^2 \int_{t_0}^t \gamma_s ds } \ket{E_0}\!\bra{E_1} +  e^{-i  \omega \delta \epsilon t} e^{- \omega^2 \delta \epsilon^2 \int_{t_0}^t \gamma_s ds} \ket{E_1}\!\bra{E_0}  \Big].
\end{align}
where $\delta \epsilon \equiv \delta E/\omega$.
The eigenvectors of $\rho_t$ are 
\begin{subequations}
\label{eq-app:eigenstates2}
    \begin{align}
\ket{\phi}  &= \frac{1}{\sqrt{2}} \left( e^{ i \omega \delta \epsilon t/2} \ket{E_0} + e^{- i \omega \delta \epsilon t/2} \ket{E_1} \right), \\
\ket{\phi_*} &= \frac{1}{\sqrt{2}} \left( e^{ i \omega \delta \epsilon t/2} \ket{E_0} - e^{- i \omega \delta \epsilon t/2} \ket{E_1} \right),
\end{align}
\end{subequations}
and the corresponding eigenvalues are
\begin{subequations}
\label{eq-app:probs2}
    \begin{align}
p &= \bra{\phi} \rho_t \ket{\phi} = 1/4 + 1/4 + 1/4 e^{-  \omega^2 \delta \epsilon^2 \int_{t_0}^t \gamma_s ds } + 1/4 e^{- \omega^2 \delta \epsilon^2 \int_{t_0}^t \gamma_s ds } = \frac{1}{2} \left( 1 + e^{- \omega^2 \delta \epsilon^2 \int_{t_0}^t \gamma_s ds} \right), \\
p_* &= \bra{\phi_*} \rho_t \ket{\phi_*} = 1/4 + 1/4 - 1/4 e^{-  \omega^2 \delta \epsilon^2 \int_{t_0}^t \gamma_s ds} - 1/4 e^{- \omega^2\delta \epsilon^2 \int_{t_0}^t \gamma_s ds} = \frac{1}{2} \left( 1 - e^{- \omega^2 \delta \epsilon^2 \int_{t_0}^t \gamma_s ds } \right).
\end{align}
\end{subequations}

The derivatives of the eigenvectors and eigenvalues with respect to $\omega$ are
\begin{subequations}
    \label{eq-app:omegaderiveigs}
    \begin{align}
        \frac{\partial }{\partial \omega} \ket{\phi}&=  i\frac{t \delta \epsilon}{2} \frac{1}{\sqrt{2}} \left( e^{ i \omega \delta \epsilon t/2} \ket{E_0} - e^{- i \omega \delta \epsilon t/2} \ket{E_1} \right) = i\frac{t \delta \epsilon}{2} \ket{\phi_*}, \\ 
        \frac{\partial }{\partial \omega} \ket{\phi_*} &=  i\frac{t \delta \epsilon}{2} \frac{1}{\sqrt{2}} \left( e^{ i \omega \delta \epsilon t/2} \ket{E_0} + e^{- i \omega \delta \epsilon t/2} \ket{E_1} \right) = i\frac{t \delta \epsilon}{2} \ket{\phi},  \\ 
        \frac{\partial }{\partial \omega} p &=  -  \omega \delta \epsilon^2 \left( \int_{t_0}^t \gamma_s ds \right) e^{- \omega^2 \delta \epsilon^2 \int_{t_0}^t \gamma_s ds},  \\ 
        \frac{\partial }{\partial \omega} p_*&=  \omega \delta \epsilon^2 \left( \int_{t_0}^t \gamma_s  ds \right) e^{- \omega^2 \delta \epsilon^2 \int_{t_0}^t \gamma_s ds}.   
    \end{align}
\end{subequations}
 
Then, using Eqs.~\eqref{eq-app:qFisher},~\eqref{eq-app:probs2}, and~\eqref{eq-app:omegaderiveigs}, the quantum Fisher information about $\omega$ is
 \begin{align}
     F_\open(\omega) &= \sum_\alpha \frac{1}{p_\alpha} \left( \frac{\partial p_\alpha}{\partial \omega} \right)^2  + 4 \sum_\alpha p_\alpha \left( \frac{\partial \bra{\alpha}}{\partial \omega} \right) \left( \frac{\partial \ket{\alpha} }{  \partial \omega} \right) - 8 \sum_{\alpha \beta} \frac{ p_\alpha p_\beta }{p_\alpha + p_\beta} \left| \bra{\beta} \frac{\partial \ket{\alpha}}{\partial \omega}\right|^2 \nonumber \\
     &=    \omega^2 \delta \epsilon^4 \left( \int_{t_0}^t \gamma_s  ds \right)^2 e^{- 2\omega^2 \delta \epsilon^2 \int_{t_0}^t \gamma_s ds} \frac{1}{p p_*}   + 4 \frac{t^2 \delta \epsilon^2}{4} - 16  p p_* \frac{t^2 \delta \epsilon^2}{4}   \nonumber \\
     &=    \omega^2 \delta \epsilon^4 \left( \int_{t_0}^t \gamma_s  ds \right)^2 e^{- 2\omega^2 \delta \epsilon^2 \int_{t_0}^t \gamma_s ds} \frac{4}{\left( 1 - e^{- \delta \epsilon^2\omega^2 \int_{t_0}^t \gamma_s ds } \right) \left( 1 + e^{- \delta \epsilon^2\omega^2 \int_{t_0}^t \gamma_s ds } \right)}   \nonumber \\
     &+  t^2 \delta \epsilon^2 - \left( 1 - e^{- \delta \epsilon^2\omega^2 \int_{t_0}^t \gamma_s ds } \right) \left( 1 + e^{- \delta \epsilon^2\omega^2 \int_{t_0}^t \gamma_s ds } \right)  t^2 \delta \epsilon^2     \nonumber \\
     &=    4 \omega^2 \delta \epsilon^4 \left( \int_{t_0}^t \gamma_s  ds \right)^2  \frac{1}{\left( e^{2 \delta \epsilon^2\omega^2 \int_{t_0}^t \gamma_s ds } - 1 \right)}   +  t^2 \delta \epsilon^2  e^{- 2 \delta \epsilon^2\omega^2 \int_{t_0}^t \gamma_s ds }.  
 \end{align}
In terms of the energy difference in the cat state, $\delta E = \omega \delta \epsilon$,
\begin{align}
\label{eq-app:FisherFrequencyCat}
    F_\open(\omega) &= 4 \frac{ (\delta E)^4 }{\omega^2} \left( \int_{t_0}^t \gamma_s  ds \right)^2  \frac{1}{\left( e^{2 (\delta E)^2 \int_{t_0}^t \gamma_s ds } - 1 \right)}   +  t^2 \frac{(\delta E)^2}{\omega^2}  e^{- 2 (\delta E)^2 \int_{t_0}^t \gamma_s ds } \nonumber \\
    &= \frac{(\delta E)^2}{\omega^2} \left(   4 (\delta E)^2    \frac{\left( \int_{t_0}^t \gamma_s  ds \right)^2}{\left( 1- e^{-2 (\delta E)^2 \int_{t_0}^t \gamma_s ds } \right)}   +  t^2   \right)  e^{- 2 (\delta E)^2 \int_{t_0}^t \gamma_s ds }
\end{align}
This proves Eq.~\eqref{eq:FfrequencyOpen} in the main text.

\subsection{A regime where noise enhances estimation precision of a global Hamiltonian parameter}

Let $\gamma_t = \gamma$ be constant and $t_0 = 0 $; i.e., the quantum sensor suffers from dephasing at a constant rate $\gamma$ throughout the whole frequency estimation protocol. Further, assume that
\begin{align}
2(\delta E)^2 \int_{t_0}^t \gamma_s ds = 2 (\delta E)^2  \gamma t = \ln(2),
\end{align}
which sets the exponential prefactor in Eq.~\eqref{eq-app:FisherFrequencyCat} to be $1/2$. Then, Eq.~\eqref{eq-app:FisherFrequencyCat} becomes 
\begin{align}
    F_\open(\omega) = \frac{(\delta E)^2}{\omega^2} \left(   4 (\delta E)^2   \gamma^2 t^2    +  \frac{1}{2} t^2   \right)  = \frac{t^2 (\delta E)^2}{\omega^2} \left(   4 (\delta E)^2   \gamma^2   +  \frac{1}{2}  \right) .
\end{align}
Thus, $F_\open(\omega)$ is larger than $F_\cl(\omega) =  (\delta E)^2 t^2/\omega^2$ whenever the following conditions hold:
\begin{subequations}
    \begin{align}
        4 (\delta E)^2   \gamma^2 \geq \frac{1}{2},   \qquad \qquad 
        t = \frac{ \ln(2) }{ 2 \gamma (\delta E)^2}.
    \end{align}
\end{subequations}

\section{An optimal estimator of time and frequency under energy decoherence}
\label{app:optimalestimators}

Consider a quantum sensor initialized in the cat state~\eqref{eq-app:Cat} suffering from energy decoherence. The sensor's state at time $t$ is
\begin{align}
\label{eq-app:stateCat}
\rho_t = \frac{1}{2} \Big[ \ket{E_0}\!\bra{E_0} + \ket{E_1}\!\bra{E_1} +  e^{i \omega \delta \epsilon t} e^{- \omega^2 \delta \epsilon^2 \int_{t_0}^t \gamma_s ds } \ket{E_0}\!\bra{E_1} +  e^{-i  \omega \delta \epsilon t} e^{- \omega^2 \delta \epsilon^2 \int_{t_0}^t \gamma_s ds} \ket{E_1}\!\bra{E_0}  \Big].
\end{align}
where $\delta \epsilon \equiv \delta E/\omega$ and $\delta E$ is the energy difference of Hamiltonian eigenstates $\ket{E_1}$ and $\ket{E_0}$.

\subsection{Qubit sensor network}

Here, I consider an $N$-qubit sensor network with a Hamiltonian $H = \omega \sum_{l = 1}^{N} \sigma_l^z/2$ to estimate time or the global field $\omega$. The sensor network starts in a GHZ state, which corresponds to $\ket{E_0} \coloneqq \otimes_j \ket{0}_j$ and  $\ket{E_1} \coloneqq \otimes_j \ket{1}_j$. The energy difference between the states is $\delta E = N \omega$.
 
Let $O \coloneqq \bigotimes_{l = 1}^N \sigma_l^x$ where $\sigma_l^x$ is the Pauli $X$ matrix of qubit $l$~\cite{GiovannettiPRL2006}. The observable's expectation value, variance, and derivatives are 
\begin{subequations}
    \begin{align}
    \label{eq-app:averageestimator}
    \langle O \rangle_t &= \tr{O \rho_t} = \frac{ e^{i \delta E t} e^{-\delta E^2 \int_{t_0}^t \gamma_s ds}  +  e^{-i  \delta E t} e^{- \delta E^2 \int_{t_0}^t \gamma_s ds} }{2} = \cos( \delta E t ) e^{-\delta E^2 \int_{t_0}^t \gamma_s ds}, \\
   (\Delta O)^2 &= 1 - \cos^2( \delta E t ) e^{- 2\delta E^2 \int_{t_0}^t \gamma_s ds}, \\
    \frac{\partial \langle O \rangle_t}{\partial t} &= - \delta E \sin ( \delta E t ) e^{-\delta E^2 \int_{t_0}^t \gamma_s ds} - \gamma_t \delta E^2 \cos( \delta E t ) e^{-\delta E^2 \int_{t_0}^t \gamma_s ds}, \\
    \frac{\partial \langle O \rangle_t}{\partial \omega} &=   - \delta \epsilon t \sin(\delta E t )  e^{-\delta E^2 \int_{t_0}^t \gamma_s ds}   - 2 \omega \delta \epsilon^2 \left( \int_{t_0}^t \gamma_s ds \right) \cos(\delta E t) e^{-\delta E^2 \int_{t_0}^t \gamma_s ds}   \nonumber \\
   &=   - \frac{\delta E t}{\omega} \sin(\delta E t )  e^{-\delta E^2 \int_{t_0}^t \gamma_s ds}   - 2 \frac{  \delta E^2 }{\omega} \left( \int_{t_0}^t \gamma_s ds \right) \cos(\delta E t) e^{-\delta E^2 \int_{t_0}^t \gamma_s ds}  ,
\end{align}
\end{subequations}
where I used that $\delta E = \omega \delta \epsilon$.

From error propagation~\cite{GiovannettiPRL2006}, the error in estimating $t$ is
\begin{align}
\label{eq-app:SaturationTimeAux}
    \var( \hat t_\opt  ) = \frac{(\Delta O)^2}{\left| \frac{\partial \langle O \rangle_t}{\partial t}  \right|^2}  &= \frac{ 1 - \cos^2( \delta E t ) e^{- 2\delta E^2 \int_{t_0}^t \gamma_s ds} }{ \left| \delta E \sin (\delta E t) + \gamma_t \delta E^2 \cos(\delta E t) \right|^2 e^{- 2 \delta E^2 \int_{t_0}^t \gamma_s ds} } .
\end{align}
The example at the end of Sec.~\ref{app:QFItime} and in the main text assumed 
\begin{align}
    \gamma_t = \dot \gamma (t-t_0), \qquad \qquad t-t_0 = \frac{\sqrt{\ln(2)}}{\sqrt{\dot \gamma} \delta E},
\end{align} 
where $\dot \gamma$ is constant and I used that $L = H$. In such a scenario, Eq.~\eqref{eq-app:SaturationTimeAux} is
\begin{align}
    \var( \hat t_\opt  ) &= \frac{ 2 - \cos^2( \delta E t )  }{ \left| \delta E \sin (\delta E t) + \gamma_t \delta E^2 \cos(\delta E t) \right|^2 }    = \frac{ 2 - \cos^2( \delta E t )  }{ \left| \delta E \sin (\delta E t) + \sqrt{\ln(2) \dot \gamma } \delta E \cos(\delta E t) \right|^2 } \nonumber \\
    &=\var_\cl(\hat t_\opt) \left( \frac{ 2 - \cos^2( \delta E t )  }{ \left| \sin (\delta E t) + \sqrt{\ln(2) \dot \gamma }  \cos(\delta E t) \right|^2 } \right).
\end{align}
This proves Eq.~\eqref{eq:CatTimeExample} in the main text.

Meanwhile, the error in estimating $\omega$ is
\begin{align}
\label{eq-app:SaturationFrequencyAux}
    \var( \hat \omega_\opt  ) = \frac{(\Delta O)^2}{\left| \frac{\partial \langle O \rangle_t}{\partial \omega}  \right|^2}  &= \frac{ 1 - \cos^2( \delta E t ) e^{- 2 \delta E^2 \int_{t_0}^t \gamma_s ds} }{ \left| \frac{\delta E}{\omega} t \sin (\delta E t) +\frac{\delta E^2}{\omega} \left( \int_{t_0}^t \gamma_s ds \right) \cos(\delta E t) \right|^2 e^{- 2 \delta E^2 \int_{t_0}^t \gamma_s ds} } .
\end{align}
The example at the end of Sec.~\ref{app:QFIfrequency} and in the main text assumed $t_0 = 0$ and
\begin{align}
    \gamma_t = \gamma, \qquad \qquad t = \frac{ \ln(2) }{ 2 \gamma (\delta E)^2}.
\end{align}
Then, using that $\var_\cl(\omega) = \frac{\omega^2}{(\delta E)^2 t^2}$, Eq.~\eqref{eq-app:SaturationFrequencyAux} becomes
\begin{align}
  \var( \hat \omega_\opt  ) &=  \frac{ 2 - \cos^2( \delta E t )   }{ \left| \frac{\delta E t}{\omega} \sin (\delta E t) +\frac{\delta E^2}{\omega} t \gamma \cos(\delta E t) \right|^2   } = \var_\cl(\omega)   \frac{ 2 - \cos^2( \delta E t )   }{ \left|  \sin (\delta E t) + \delta E \gamma \cos(\delta E t) \right|^2   }.
\end{align}
When $\delta E \gamma$ is an integer multiple of $\pi$, the quotient $\var( \hat \omega_\opt  )/\var_\cl(\omega)$ decreases as $1/(\gamma \delta E)^2$, as does the analytical expression for $F_\cl(\omega)/F_\open$ in Sec.~\ref{app:QFIfrequency}. Thus, $\langle O \rangle$ yields an optimal estimator of $\omega$ that benefits from the incoherent noise considered.

\subsection{Photonic quantum sensor}
 
Consider a two-mode photonic sensor with 
a Hamiltonian $H = \omega a^\dag a /2 + \omega b^\dag b /2$. The sensor is initialized in a NOON state, where $\ket{E_1} = \ket{0}_A\otimes\ket{N}_B \equiv \ket{0,N}$ and $ \ket{E_0} = \ket{N}_A\otimes\ket{0}_B \equiv \ket{N,0}$ are eigenvectors of the Hamiltonian~\cite{NOON1, NOON2}. Affected by energy decoherence at a rate $\gamma_t$, the sensor's state at time $t$ is given by Eq.~\eqref{eq-app:stateCat}, where $\delta E = N \omega$.

One can use the observable $O \coloneqq \ket{0,N}\!\bra{N,0} + \ket{N,0}\!\bra{0,N}$ as an estimator of $t$ and $\omega$~\cite{Lee_2002_NOON, mitchell2004superNOON}. The observable's expectation value is
\begin{align}
    \langle O \rangle_t &= \tr{O \rho_t} = \frac{ e^{i \delta E t} e^{-\delta E^2 \int_{t_0}^t \gamma_s ds}  +  e^{-i  \delta E t} e^{- \delta E^2 \int_{t_0}^t \gamma_s ds} }{2} = \cos( \delta E t ) e^{-\delta E^2 \int_{t_0}^t \gamma_s ds},
\end{align}
identical to Eq.~\eqref{eq-app:averageestimator}. Thus, the calculations that follow Eq.~\eqref{eq-app:averageestimator} hold, so $O$ serves as an optimal estimator that benefits from incoherent noise in a photonic quantum sensor.

\end{document}